\documentclass[preprint]{aastex}

\newcommand\kms           {km~s$^{-1}$}
\newcommand\lsr           {\ensuremath{_{\mathrm {LSR}}}}

\newcommand\myhrulefill   {\leavevmode \leaders \hrule height3pt depth-2.6pt \hfill
 \kern 0pt }
\newcommand{\rarr}        {\ensuremath{\rightarrow}}
\newcommand{\tnm}         {\tablenotemark}
\newcommand{\tnt}         {\tablenotetext}

\shorttitle{Interstellar Hydroxyl Masers.~II.}
\shortauthors{Fish et al.}

\begin{document}

\title{Interstellar Hydroxyl Masers in the Galaxy.~II.~Zeeman Pairs and the
Galactic Magnetic Field}
\author{Vincent~L.~Fish, Mark~J.~Reid and Alice~L.~Argon}
\affil{Harvard--Smithsonian Center for Astrophysics}
\affil{60 Garden Street, Cambridge, MA  02138}
\email{vfish@cfa.harvard.edu, mreid@cfa.harvard.edu, aargon@cfa.harvard.edu}
\and
\author{Karl~M.~Menten}
\affil{Max-Planck-Institut f\"{u}r Radioastronomie}
\affil{Auf dem H\"{u}gel 69, Bonn, D-53121  Germany}
\email{kmenten@mpifr-bonn.mpg.de}

\begin{abstract}
We have identified and classified Zeeman pairs in the survey by Argon,
Reid, \& Menten of massive star-forming regions with 18~cm
($^{2}\Pi_{3/2}, J = 3/2$) OH maser emission.  We have found a total
of more than 100 Zeeman pairs in more than 50 massive star-forming
regions.  The magnetic field deduced from the Zeeman splitting has
allowed us to assign an overall line-of-sight magnetic field direction
to many of the massive star-forming regions.  Combining these data
with other data sets obtained from OH Zeeman splitting, we have looked
for correlations of magnetic field directions between star-forming
regions scattered throughout the Galaxy.  Our data do not support a
uniform, Galactic-scale field direction, nor do we find any strong
evidence of magnetic field correlations within spiral arms.  However,
our data suggest that in the Solar neighborhood the magnetic field
outside the Solar circle is oriented clockwise as viewed from the
North Galactic Pole, while inside the Solar circle it is oriented
counterclockwise.  This pattern, including the magnetic field reversal
near the Sun, is in agreement with results obtained from pulsar
rotation measures.
\end{abstract}


\section{Introduction}

Hydroxyl (OH) masers are an excellent tool for probing the environment
of massive star-forming regions (SFRs).  They are bright and easily
observable and act as point tracers of the local velocity and magnetic
fields, owing to their large Zeeman splitting coefficient.  For the
ground-state $^{2}\Pi_{3/2}, J = 3/2$ transitions, the Zeeman
splitting coefficient is as large as 0.59 km~s$^{-1}$~mG$^{-1}$.  For
a typical maser linewidth of 0.3 km~s$^{-1}$, a milligauss-strength
magnetic field is sufficient to cleanly separate Zeeman components of
the 1665 MHz transition, and even fields as weak as several hundred
microgauss can be detected.  In addition, the sense of the splitting
indicates the projected line-of-sight magnetic field orientation.

\citet{davies} first hypothesized that a connection exists between
magnetic field directions obtained from molecular Zeeman splitting and
a large-scale Galactic magnetic field.  He found that eight OH maser
sources (and several \ion{H}{1} clouds) mostly near the Solar circle
with observable Zeeman splitting displayed magnetic fields aligned in
the clockwise direction when viewed from the North Galactic Pole.
\citet{reidsilver} conducted a literature search and found a total of
17 reliable OH maser sources with detectable Zeeman pairs in the
$^2\Pi_{3/2}, J = 3/2$ transitions.  Of these 17, 14 had line-of-sight
field directions consistent with an overall clockwise Galactic field.
In the largest survey to date, \citet{baudry} extended this dataset to
include a total of 46 sources with Zeeman pairs in the $^2\Pi_{3/2}, J
= 3/2$ and $J = 5/2$ transitions of OH.  Excluding the Galactic
center, they found 28 sources consistent with a clockwise magnetic
field and 17 sources consistent with a counterclockwise field.

\citet[henceforth Paper I]{arm} conducted a survey of 91 interstellar
regions with known $^2\Pi_{3/2}, J = 3/2$ OH maser emission.  They
identified and mapped maser spots in the regions.  In this paper we
extend the analysis of Paper I by identifying Zeeman pairs toward
massive SFRs, many of which contain
ultracompact (UC) \ion{H}{2} regions.
%\footnote{Many interstellar OH
%masers are located near detectable UC\ion{H}{2} regions.  However,
%some interstellar OH masers may form while the precursors of
%UC\ion{H}{2} regions are still contained by rapid accretion
%\citep{keto, churchwell} and are not yet detectable.  Throughout this
%paper, we use the term ``UC\ion{H}{2}'' to refer to the central
%condensation of massive star-forming regions, even if a UC\ion{H}{2}
%region is not (yet) detectable.}
We also investigate whether or not
the magnetic fields deduced from OH Zeeman splitting around
massive SFRs are related to the large-scale magnetic field of
the Galaxy.

\section{Observations}

A total of 91 maser sources were observed by Paper I with the A
configuration of the National Radio Astronomy Observatory's Very Large
Array\footnote{The National Radio Astronomy Observatory is a facility
of the National Science Foundation operated under cooperative
agreement by Associated Universities, Inc.} in Socorro, NM.  All
sources were observed in both left- and right-circular polarization in
at least one main-line $^2\Pi_{3/2}, J = 3/2$ 18~cm OH transition
(1665.4018 and 1667.3590~MHz), and some were also observed in one or
both of the satellite-line transitions (1612.2310 and 1720.5300~MHz).
The 0.763~kHz spectral channel separation used corresponds to a
velocity separation of 0.14~\kms\ for the main-line transitions (FWHM
$\sim 20\%$ larger).  Further details of the observations can be found
in Paper I.

\section{Zeeman Pair Selection Criteria\label{criteria}}

In the presence of a magnetic field, the degeneracy of the magnetic
sublevels of the OH molecule is broken.  For any given hyperfine
transition, this results in three or more spectral lines with
differing polarization properties --- the Zeeman effect
\citep[see][for more details]{davies}.  OH masers often display pairs
of strong left and right elliptically polarized $\sigma$-components.
For reasons that are not well understood, purely linearly polarized
$\pi$-components are rarely, if ever, detected.

The task of identifiying Zeeman pairs --- two oppositely elliptically
polarized lines at the same position on the sky --- is relatively
simple, provided one has sufficient angular resolution to resolve
maser clusters \citep[e.g.,][]{reidw3}.  However, with the VLA one can
only achieve an angular resolution of $\approx 1$~arcsec for the OH
transitions at wavelengths of 18~cm.  Were the emission in any
spectral channel to come from only one (unresolved) spot on the sky,
this would not be a significant limitation.  However, in reality the
emission in any given spectral channel can come from spectral blends
of masers at different positions in the source separated by less than
the beam size.  Thus, great care must be taken to determine Zeeman
pairs from the Paper I data.

In order to identify Zeeman pairs, a modified version of the NRAO
Astronomical Image Processing System (AIPS) task JMFIT was used to
identify the brightest maser emission spot in each spectral channel
(velocity plane) for any given source.  A maser line was identified
when emission was seen from approximately the same position in at
least three adjacent velocity channels, with the strongest emission
not found on the edge channels.  This resulted in a large data set
that could be automatically searched for Zeeman pairs.

Ideally, one would like to find Zeeman pairs in which the oppositely
circularly polarized components were coincident on the sky to better
than their apparent spot sizes.  However, \citet{reidw3} showed that
OH masers in W3(OH) clustered on a scale of $3\times 10^{15}$~cm
($0\arcsec\!\!.09$ at 2.2~kpc; distances are discussed in \S
\ref{distances} and listed in Table \ref{pairs_tab}) and that
oppositely circularly polarized maser lines within a cluster gave
reliable magnetic field measurements when interpreted as Zeeman pairs.
Thus, we searched for oppositely circularly polarized maser lines with
projected positions separated by less than $3\times 10^{15}$~cm.
While the main-line 1.6~GHz $\sigma$-components are in general
\emph{elliptically} polarized, it is almost functionally equivalent to
treat them as being \emph{circularly} polarized for the purposes of
right-circular polarization (RCP) and left-circular polarization (LCP)
VLA data.  The caveat is that, for example, a right-elliptically
polarized maser component, being the sum of circularly and linearly
polarized emission, may appear in \emph{both} RCP and LCP data at the
same velocity, although the LCP emission will be much weaker, as the
LCP feed is sensitive to only half of the (presumed small) linear
component of the full elliptical polarization.

Some Zeeman pairs consisted of single RCP and LCP lines that were
coincident on the sky to better than $3\times 10^{15}$~cm.  We
consider these easily recognizable and reliable Zeeman pairs.
However, more than one maser line in the same polarization would often
be seen within the assumed $3 \times 10^{15}$~cm clustering length,
such that it was unclear which should be paired with the maser spot in
the other polarization.  While such data are insufficient to make an
unambiguous Zeeman pair identification in all such cases, sometimes
corroborating evidence from another maser transition (or occasionally
even the same transition) in the same spatial region allowed the
identification of Zeeman pairs.  In other cases, the line-of-sight
field direction was deducible, even if unambiguous identification of a
Zeeman pair was not possible, for example, when multiple LCP and RCP
lines were seen, but each RCP line was at a higher velocity than any
of the LCP lines.  We felt it important to generate specific,
objective, criteria that allowed an algorithmic approach to search for
Zeeman pairs and accept or reject them without resorting to a visual
examination of spectra on a case-by-case basis, which can lead to
significant unquantifiable biases.  Below we describe how we handled
each of the complications.

For any given source, a ``window'' of radius $3 \times 10^{15}$~cm was
centered on the strongest line in either polarization in a single
transition.  In principle, if the window included one line in each
polarization, the lines would be identified as a Zeeman pair.
However, emission channels key to identifying the second maser line in
a Zeeman pair sometimes lay just outside the window when it was
centered on the position of brightest emission, whereas were the
window moved between the position of the lines, it would pass the $< 3
\times 10^{15}$~cm separation criterion.  To overcome this problem,
the window center was moved from the centroid of strongest emission
along a grid, in increments of 5\% of the window radius, out to a
maximum displacement of 50\% of the radius in both the R.A. and Dec
directions.  The window containing the largest number of lines was
used, provided that at least one line of each polarization was
included.  ``Tiebreaker'' criteria consisted first of choosing the
window with the largest number of velocity channels represented and
second of choosing the window whose offset from the grid center was
smallest.  Next, all velocity channels contained in the chosen window
were flagged as used, and the process was iterated with the remaining
maser spots.  In this way, the entire source was partitioned into
windows consisting of potential Zeeman pairs.  This partitioning was
done independently for each maser transition for which data were
collected for the source.

Identification of Zeeman pairs in windows containing only one maser
line in each polarization is straightforward.  Figure 1a shows spectra
from two windows for G351.775$-$0.538, one in each of the 1665 and
1667~MHz transitions.  The 1665~MHz window contains two oppositely
circularly-polarized maser lines.  Removing the effects of the Zeeman
splitting from a magnetic field of $-3.80 \pm 0.03$~mG (i.e., oriented
in the hemisphere toward the Sun.), the center velocity of the
emission would be at $-26.74 \pm 0.02$~\kms.  The 1667~MHz window,
centered 120~mas away, shows a similar splitting of $-3.66 \pm
0.06$~mG centered at a velocity of $-26.79 \pm 0.02$~\kms.  While a
large fraction of the Zeeman pairs we identified were found in windows
containing exactly one maser line in each polarization, a few of these
pairs appeared with a corroborating nearby pair of a similar magnetic
field in another transition.

Table \ref{pairs_tab} is a list of Zeeman pairs with ``non-zero''
velocity separation.  The flux, velocity, and velocity width of lines
was determined by a Gaussian fit to the center 5 channels of each line
when available, else to 4 or only 3 channels.  (By our criteria for a
maser line, the center 3 channels must necessarily be detected.)
Pairs are only included in this table if the velocity difference
between the LCP and RCP lines exceeds twice the
signal-to-noise-limited velocity error, taken to be the sum in
quadrature of the HWHM line width divided by the signal-to-noise ratio
of each of the lines.  However, in no case is a pair of lines allowed
to satisfy this criterion if the velocity center separation is less
than 5\% of the quadrature sum of the line widths, or typically about
0.02~\kms.  These ``pairs'' may be actual Zeeman pairs with a small
splitting due to a small magnetic field, or they may be single maser
spots with a large linear polarization fraction, but the limited
spatial and velocity resolution of the VLA and the absence of
full-polarization observations render it difficult to decide between
these two cases.  These ``near zero'' velocity separation maser spots
are listed in Table 2.

While identification of many of the pairs listed in Table 1 was
straightforward, not all windows were so simple to interpret.  When a
window contains more than one maser line in a polarization,
interpreting the Zeeman pattern is more complicated.  In this case, it
may still be possible to know the light-of-sight direction of the
magnetic field, but not its strength.  Figure 1b shows the spectrum of
one window associated with G032.744$-$0.076.  There is one RCP maser
line (labelled 1) centered at 32.38~\kms\ and two LCP lines (2 and 3)
at higher velocities (34.60 and 35.02~\kms, respectively).
Additionally, there is one feature (4) that does not meet our criteria
as a maser line, since the low-velocity side of the line is not seen,
probably owing to blending with a stronger line elsewhere in the
source.  It is unclear which LCP line should be associated with the
RCP line as a Zeeman pair.  Regardless of the exact pairing, the RCP
line is at lower velocity than any LCP line, so the line-of-sight
orientation of the magnetic field is known to be pointed toward the
Sun.  Cases such as this, where the field direction is known but its
strength is not, are frequent among our data.  When a range of values
is quoted for the magnetic field in Table 1, the direction of the
magnetic field is well-determined, and the smallest and largest
possible magnetic field intensities are given.  Some line parameters
not listed in the table are given in the notes on individual sources.

%In some cases, a very strong line ($\gtrsim$100
%Jy) in one polarization appeared with a strong line and a weak line in
%the opposite polarization.  When the flux ratio between lines of
%opposite polarization exceeded 5, we dropped the weakest line and
%reconsidered the pairing in the window, often resulting in a Zeeman
%pair with a definite magnetic field intensity.  While this flux ratio
%is somewhat arbitrary, it reflects our empirical belief that strong
%maser lines tend to be accompanied by a fairly strong line in the
%opposite polarization.

Sometimes when a window contains enough maser lines such that the
magnetic field strength or direction is ambiguous, supporting evidence
may exist to suggest one pairing above all others.  If a pairing in
one transition implies a magnetic field strength and direction
consistent with that of a pair in another transition at nearly the
same location, that pairing is accepted in preference to all other
possible pairings.  Figure 1c shows a 1665~MHz and a 1667~MHz window
in K3$-$50, coincident to within 46~mas.  In each window, there is one
RCP line (1) at lower velocity than either of the two LCP lines (2 and
3), so the line-of-sight field is oriented toward the Sun.  But the
field magnitude can also be deduced, assuming that the nearly
overlapping windows in the two transitions are measuring the field at
approximately the same location.  In the 1665~MHz window, pairing
lines 1 and 2 would give a magnetic field of $-1.5$~mG, while the 1--3
pairing implies a $-2.5$~mG field.  In the 1667~MHz window, the 1--2
pairing yields $-0.8$~mG, while 1--3 yields $-2.6$~mG.  The close
agreement between the field values for the 1--3 pairs for the two
transitions suggests that these Zeeman pairs should be favored over
the other pairing possibilities.  For definiteness, when appealing to
corroborating evidence from one transition to resolve pairing
ambiguities in another transition, we required $\Delta B \leq 0.4$~mG,
$\Delta v_c \leq 0.2$~\kms, and $\Delta\theta \leq 0\arcsec\!\!.2 \sec
z \tan z$, where $\Delta B$ is the difference of the implied magnetic
field strengths, $\Delta v_c$ is the separation of the center velocity
of each pair, $\Delta\theta$ is the positional offset between pairs,
and $z$ is the zenith angle of the source as viewed from the location
of the VLA at transit.  As noted in Paper I, relative positional
offsets between different OH transition frequencies are $\approx
0\arcsec\!\!.1$ for high-declination sources, but grow to several
arcseconds for sources at very low declination ($\lesssim -40\degr$).
The function ``$\sec z \tan z$'' was chosen in the spirit of the
plane-parallel approximation for the atmospheric contribution to the
relative error between observations of different transitions
\citep[see][]{reid99}.

If two pairings in the same window in the same transition imply the
same magnetic field direction and strength, those pairings are also
accepted over all other pairings.  This situation arises when two
Zeeman pairs in the same transition are encompassed in the same window.
Figure 1d shows a 1665~MHz window for G45.465$+$0.047 that contains a
cluster of maser lines.  Since all of the RCP lines are at a higher
velocity than the LCP lines, the magnetic field is oriented away from
the Sun.  However, the velocities associated with the line centers
favor two particular pairings: pairing lines 1 and 4 yields a magnetic
field strength of 3.7~mG, and 2 and 5 yield 3.8~mG.

Occasionally, a window would produce a spectrum with a clear, strong
line in one polarization and an incomplete line in the opposite
polarization.  A closer look at the data would show that one or two
critical points of the incomplete line lay just outside the window.
By increasing the size of the window, a Zeeman pair was often
identified.  Using a larger window (generally twice the clustering
radius) can be justified for two reasons.  First, the window radius of
$3 \times 10^{15}$~cm is based on the clustering scale analysis of
\citet{reidw3}, which is itself based on only one source: W3(OH),
assuming the photometric distance of 2.2~kpc \citep{humphreys}.  The
kinematic distance of 4.3~kpc, assuming the NH$_3$ LSR velocity of
$-45$~\kms\ \citep{rmb} and a flat $R_0 = 8.5$~kpc, $\Theta(R) =
220$~\kms\ rotation curve, would correspond to a clustering scale of
$6 \times 10^{15}$~cm, or twice that which is derived from the
photometric distance.  In any case, it is unclear to what extent this
varies among massive SFRs.  Second, the distance to some of
the massive SFRs is unknown.  If the assumed distance to the
source is greater than the actual distance, the (angular) window
radius used would be proportionally smaller than the desired $3 \times
10^{15}$~cm.  For sources for which the kinematic distance ambiguity
has not been resolved, the Zeeman pair identification analysis was
done twice, assuming each of the near and far distances for
calculation of the clustering scale.  Zeeman pairs identified assuming
either distance were accepted and are listed in Table \ref{pairs_tab}.
Despite the distance uncertainty, we believe that Zeeman pairs
identified around these sources are still reliable indicators of the
local magnetic field.  When the far kinematic distance is more than a
few times the near kinematic distance, possible errors involve
partitioning a far source into windows that are too large or a near
source into windows that are too small.  In the former case, nearly
all of the maser emission of a source (presumedly) located at the far
kinematic distance would be concentrated in a single window when the
near distance was used for the angular clustering scale, resulting in
confusion of Zeeman pairs.  In the latter case, a (presumed) near
source would be micropartitioned into windows of small radius,
rejecting many Zeeman pairs coincident to within the desired linear
clustering scale.  Thus, misresolving the near/far kinematic distance
ambiguity results in the identification of fewer Zeeman pairs,
regardless of whether a far source is assumed to be near or a near
source is assumed to be far.

Zeeman pairs were also graded to reflect the uncertainty in pairing.
Pairs from windows containing only one line in each polarization were
assigned a grade of B, unless they were confirmed by a similar pair in
a second transition, in which case they were given a grade of A.  All
pairs whose pairing was ambiguous were given a grade of C, even if the
ambiguity was resolved by appealing to a second pair in the same or
another transition.  Pairs for which the separation of the Zeeman
components exceeded the default window size were assessed a grade of
D.  In one case in G081.871+0.781, a grade of D was assigned to a
Zeeman pair believed to be spurious; for further details see \S
\ref{sourcenotes}.  This grading system is in the spirit of, although
not identical to, the system used in \citet{reidsilver}, in which the
main emphasis was on whether interferometry was used to identify
Zeeman splitting, and if so, at what resolution.  All of the Zeeman
pairs listed in Table \ref{pairs_tab} were identified using
interferometry.

\begin{deluxetable}{llrcrrcrrccr@{}lc}
\tabletypesize{\scriptsize}
\rotate
\tablewidth{601.9319pt}
\tablecaption{Zeeman Pairs\label{pairs_tab}}
\tablehead{
  &&&&\multicolumn{3}{c}{\myhrulefill LCP \myhrulefill}&\multicolumn{3}{c}{\myhrulefill RCP \myhrulefill}&&&\\
  \colhead{\phm{Source}} &  \colhead{\phm{Alias}} &   
  \colhead{Dist.\tnm{a}} & \colhead{Freq.} &
  \colhead{$S$\tnm{b}} &  \colhead{$v\lsr$\tnm{b}} & \colhead{$\Delta v$\tnm{b}} &
  \colhead{$S$\tnm{b}} &  \colhead{$v\lsr$\tnm{b}} & \colhead{$\Delta v$\tnm{b}} &
  \colhead{$\Delta\theta$\tnm{c}} &
  \multicolumn{2}{c}{$B$\tnm{d}} &      \colhead{\phm{Grade}}    \\
  \colhead{Source} &   \colhead{Alias} &   \colhead{(kpc)} &
  \colhead{(MHz)} &
  \colhead{(Jy)} &     \colhead{(\kms)} &  \colhead{(\kms)} &
  \colhead{(Jy)} &     \colhead{(\kms)} &  \colhead{(\kms)} &
  \colhead{(arcsec)} & 
  \multicolumn{2}{c}{(mG)} &     \colhead{Grade\tnm{e}} 
}
\startdata
G000.547$-$0.852 & RCW 142     & 7.3 & 1667 &   0.23 &    8.96 & 0.65 &   3.72 &    6.33 & 0.43 & 0.046 &  --7&.4 & B \\
G000.666$-$0.034 & Sgr B2M     & 8.2 & 1720 &   6.84 &   61.09 & 0.28 &  14.21 &   61.25 & 0.32 & 0.028 &  0&.7 & B\\
G000.670$-$0.058 & Sgr B2      & 8.2 & 1612 &   3.78 &   67.40 & 0.87 &   3.91 &   67.24 & 0.82 & 0.024 &  --0&.7 & D\\*
G005.886$-$0.393 &\nodata      & 3.8 & 1665 &   4.26 &    9.52 & 0.26 &   3.04 &   10.41 & 0.32 & 0.028 &   1&.5 & A\\*
                 &             &     & 1667 &   8.90 &    9.71 & 0.28 &   7.83 &   10.22 & 0.29 & 0.012 &   1&.4 & A\\*
                 &             &     &      &   1.43 &    4.93 & 0.75 &   0.74 &    5.18 & 0.96 & 0.042 &   0&.7 & B\\
                 &             &     &      &   1.07 &    6.10 & 0.80 &   1.50 &    7.08 & 0.56 & 0.080 &  0.2&\rarr 2.8 &C \\
G009.622$+$0.195 &\nodata      & 5.7 & 1665 &   2.41 &  --1.07 & 0.42 &   8.06 &    1.75 & 0.37 & 0.018 &  3.9&\rarr  9.7 &C \\*
                 &             &     &      &   1.34 &    3.90 & 0.53 &   2.68 &    3.85 & 0.40 & 0.044 & --0.1&\rarr --5.4 &C \\*
                 &             &     & 1667 &   9.67 &    1.47 & 0.41 &   1.98 &    1.43 & 0.48 & 0.043 &  --0&.1 & B\\
G010.624$-$0.385 &\nodata      & 4.8 & 1667 &  43.03 &  --0.59 & 0.35 &  44.13 &  --1.98 & 0.25 & 0.045 &  --3&.9 & B\\ 
                 &             &     &      &   0.80 &    2.08 & 0.30 &   1.26 &    0.10 & 0.33 & 0.049 &  --5&.6 &B \\
G012.216$-$0.117 &\nodata      & 3.3 & 1665 &  15.26 &   29.07 & 0.49 &  12.32 &   27.55 & 0.50 & 0.042 &  --2&.6 &B \\
G012.680$-$0.181 & W33 B       & 4.0 & 1665 &  20.29 &   64.45 & 0.50 &   4.45 &   63.34 & 0.43 & 0.035 & --1.9&\rarr --6.3 &C \\*
                 &             &     & 1667 &   5.45 &   64.50 & 0.48 &   3.62 &   62.50 & 0.62 & 0.029 & --0.7&\rarr --7.5 &C \\
G012.908$-$0.259 & W33 A       & 4.0 & 1667 &  69.34 &   39.93 & 0.49 &   5.12 &   40.41 & 0.75 & 0.023 &   1&.4 &B \\
G017.639$+$0.155 &\nodata      & 2.1 & 1665 &   1.79 &   20.92 & 0.35 &   2.22 &   20.42 & 0.28 & 0.048 &  --0&.8 &B \\
G020.081$-$0.135 &\nodata      &12.3 & 1665 &   0.66 &   42.90 & 0.48 &   2.22 &   48.64 & 0.39 & 0.060 &   9&.7 &B \\*
                 &             &     &      &  10.99 &   46.60 & 0.39 &   1.42 &   49.75 & 0.48 & 0.016 &  5.3&\rarr 7.3 &C \\
G028.199$-$0.048 &\nodata      & 5.7 & 1665 &  11.48 &   94.92 & 0.36 &  19.23 &   94.89 & 0.27 & 0.008 &  --0&.1 &B \\
G031.412$+$0.307 &\nodata      & 6.2 & 1667 &   3.50 &  103.84 & 0.57 &   2.80 &  103.48 & 0.60 & 0.013 &  --1&.0 &B \\
G032.744$-$0.076 &\nodata      & 2.3 & 1665 &   0.61 &   32.84 & 0.67 &   0.68 &   31.00 & 0.56 & 0.099 &  --3&.1 &B \\*
                 &             &     &      &   1.04 &   35.02 & 0.58 &   1.73 &   32.38 & 0.46 & 0.059 & --3.8&\rarr --4.5 &C \\*
                 &             &     &      &   0.39 &   36.44 & 0.37 &   0.92 &   39.58 & 0.35 & 0.053 &  5.3&\rarr 6.4 &C \\
G034.257$+$0.154 &\nodata      & 3.8 & 1665 &  16.04 &   58.14 & 0.37 &  79.02 &   58.22 & 0.38 & 0.015 &   0&.1 &B \\*
                 &             &     & 1667 & 108.48 &   58.64 & 0.38 &  66.28 &   58.49 & 0.37 & 0.005 &  --0&.4 &B \\*
                 &             &     &      &   7.28 &   57.50 & 0.32 &  10.52 &   57.41 & 0.28 & 0.006 & --0.3&\rarr --1.6 &C \\
G035.024$+$0.350 &\nodata      & 3.1 & 1665 &   2.74 &   44.62 & 0.24 &   8.00 &   47.07 & 0.47 & 0.028 &  0.0&\rarr 5.5 &C \\
G035.197$-$0.743 &\nodata      & 2.2 & 1665 &   1.91 &   26.71 & 0.36 &   5.42 &   28.82 & 0.32 & 0.036 &  3.6&\rarr  4.2 &C \\
G035.577$-$0.029 &\nodata      &10.5 & 1665 &  29.48 &   48.94 & 0.42 &   2.86 &   48.25 & 0.38 & 0.103 &  --1&.2 &D \\*
                 &             &     &      &   6.87 &   50.79 & 0.33 &  57.64 &   49.01 & 0.28 & 0.116 & --1.5&\rarr --4.9 &D \\
G040.622$-$0.137 &\nodata      & 2.1 & 1665 &  96.06 &   32.57 & 0.31 &  37.90 &   32.62 & 0.40 & 0.010 &   0&.1 &B \\*
                 &             &     &      &   1.31 &   34.47 & 0.34 &   2.14 &   35.74 & 0.32 & 0.020 &  0.5&\rarr 2.2 &C \\*
                 &             &     & 1667 &   0.22 &   34.74 & 0.33 &   1.02 &   31.67 & 0.46 & 0.033 &  --8&.7 &B \\*
                 &             &     &      &  18.79 &   31.97 & 0.32 &   1.89 &   30.74 & 0.32 & 0.097 & --0.3&\rarr --3.5 &C \\
G043.148$+$0.015 & W49         &11.5 & 1612 &   2.30 &   13.20 & 0.26 &   2.59 &   13.88 & 0.44 & 0.037 &   2&.9 &D \\
G043.165$-$0.028 & W49 S       &11.5 & 1665 &  63.22 &   15.33 & 0.44 & 221.62 &   16.22 & 0.34 & 0.008 &   1&.5 &B \\*
                 &             &     &      &  45.65 &   19.11 & 0.36 &   6.60 &   22.55 & 0.52 & 0.016 &   5&.8 &B \\*
                 &             &     &      &  26.02 &   13.23 & 1.06 &  35.52 &   13.68 & 0.57 & 0.013 &   0&.8 &B \\*
G043.796$-$0.127 &\nodata      & 3.0 & 1665 &  95.57 &   41.53 & 0.40 &  76.30 &   41.59 & 0.45 & 0.024 &   0&.1 &B \\
G045.071$+$0.134 &\nodata      & 4.7 & 1665 &  13.69 &   53.94 & 0.39 &   9.17 &   54.21 & 0.28 & 0.008 & 0.5&\rarr 5.0 &C \\*
                 &             &     & 1667 &   6.18 &   53.53 & 0.28 &   2.16 &   55.26 & 0.30 & 0.033 &  2.9&\rarr 4.9 &C \\*
                 &             &     &      &   5.60 &   55.32 & 0.34 &   3.54 &   56.33 & 0.55 & 0.008 &  1.6&\rarr 5.2 &C \\
G045.122$+$0.133 &\nodata      & 6.0 & 1667 &   1.90 &   52.98 & 0.38 &   2.08 &   52.18 & 0.36 & 0.005 &  --2&.3 &B \\
G045.465$+$0.047 &\nodata      & 5.8 & 1665 &  18.26 &   65.97 & 0.41 &   4.32 &   68.18 & 0.54 & 0.011 &   3&.7 &C \\*
                 &             &     & 1665 &  14.51 &   65.34 & 0.47 &   1.14 &   67.52 & 0.52 & 0.033 &   3&.7 &C \\
G049.469$-$0.370 & W51         & 7.3 & 1720 &   1.06 &   64.01 & 0.55 &   0.70 &   64.80 & 0.55 & 0.021 &   3&.3 &B \\
G049.488$-$0.387 & W51 M/S     & 6.5 & 1665 &  70.41 &   59.24 & 0.49 & 164.62 &   58.25 & 0.57 & 0.032 & --1.1&\rarr --1.7 &C \\*
                 &             &     & 1720 &  38.30 &   56.97 & 0.53 &  87.99 &   58.16 & 0.44 & 0.001 &   5&.0 &B \\*
                 &             &     &      &  25.74 &   58.60 & 0.50 &  61.24 &   59.55 & 0.72 & 0.001 &   4&.0 &B \\*
                 &             &     &      &   5.87 &   55.76 & 0.46 &   7.94 &   56.47 & 0.40 & 0.013 &   3&.0 &B \\*
                 &             &     &      &   1.96 &   52.64 & 0.31 &   2.42 &   53.47 & 0.44 & 0.013 &   3&.5 &B \\
G069.540$-$0.976 & ON 1        & 1.3 & 1665 &   4.50 &   13.97 & 0.50 &  15.48 &   11.86 & 0.27 & 0.127 &  --3&.6 &B \\*
                 &             &     &      &   0.64 &   10.87 & 0.29 &   0.84 &   10.74 & 0.43 & 0.074 &  --0&.2 &B \\*
                 &             &     &      &  10.21 &   13.24 & 0.40 &  22.65 &   13.12 & 0.41 & 0.030 & --0.2&\rarr --3.9 &C \\
G070.293$+$1.601 & K3--50      & 8.0 & 1665 &   2.37 & --21.28 & 0.93 &   2.57 & --22.78 & 2.30 & 0.006 &  --2&.5 &B \\*
                 &             &     &      &  11.52 & --19.76 & 0.26 &   5.64 & --21.27 & 0.46 & 0.008 & --2&.5 &C \\*
                 &             &     & 1667 &   3.52 & --20.15 & 0.35 &   3.45 & --21.07 & 0.41 & 0.010 & --2&.6 &C \\
G081.721$+$0.571 & W75 S       & 2.0 & 1665 &   7.61 &  --1.29 & 0.27 &   1.40 &  --4.00 & 0.28 & 0.102 &  --4&.6 &B \\*
                 &             &     &      &  21.58 &    1.31 & 0.35 &   3.90 &    5.19 & 0.27 & 0.023 &  2.8&\rarr 9.3 &C \\
G081.871$+$0.781 & W75 N       & 2.0 & 1665 &  14.61 &    9.35 & 0.20 &  48.50 &   12.47 & 0.30 & 0.007 &   5&.3 &B \\*
                 &             &     &      &  16.92 &    5.16 & 0.38 &   5.32 &    5.23 & 0.36 & 0.022 &   0&.1 &B \\*
                 &             &     & 1667 &  26.82 &    9.34 & 0.36 &  18.80 &    9.28 & 0.32 & 0.015 &  --0&.2 &B \\*
                 &             &     &      &   0.72 &   10.05 & 0.39 &   0.70 &    1.06 & 0.55 & 0.037 & --25&.4 &D\tnm{f}\\*
                 &             &     &      &   1.84 &    5.44 & 0.34 &   0.97 &    8.10 & 0.12 & 0.104 &  7.5&\rarr 11.3 &C \\
G109.871$+$2.114 & Cep A       & 0.7 & 1665 &   0.29 &  --5.05 & 0.62 &   0.29 &  --7.10 & 0.82 & 0.135 &  --3&.5 &B \\*
                 &             &     &      &   9.31 & --16.22 & 0.28 &   7.28 & --14.23 & 0.28 & 0.006 &   3&.4 &C \\*
                 &             &     & 1667 &   3.78 & --15.77 & 0.37 &   4.34 & --14.64 & 0.30 & 0.012 &   3&.2 &B \\
G111.533$+$0.757 & NGC 7538    & 2.8 & 1665 &   2.78 & --54.33 & 0.33 &   2.38 & --52.99 & 0.33 & 0.005 &   2&.3 &B \\
G133.946$+$1.064 & W3 OH       & 2.2 & 1612 &   1.30 & --43.77 & 0.54 &   6.27 & --42.20 & 0.45 & 0.033 &  3.6&\rarr 6.7 &C \\*
                 &             &     & 1667 &  25.17 & --44.45 & 0.58 &  21.27 & --42.19 & 0.31 & 0.104 &   6&.4 &B \\*
                 &             &     &      &   1.92 & --47.76 & 0.36 &   0.60 & --47.82 & 0.26 & 0.026 &  --0&.2 &B \\*
                 &             &     & 1720 &   0.41 & --44.38 & 0.34 &   5.29 & --43.12 & 0.27 & 0.066 &   5&.3 &B \\*
                 &             &     &      &  10.39 & --45.56 & 0.49 &   7.55 & --44.72 & 0.66 & 0.024 &  2.2&\rarr 12.0 &C \\
G213.706$-$12.60 & Mon R2      & 0.9 & 1665 &   0.12 &   12.36 & 0.46 &   0.62 &   10.37 & 0.38 & 0.067 &  --3&.4 &B \\
G341.219$-$0.212 &\nodata      & 3.2 & 1665 &   3.49 & --40.85 & 0.45 &  10.06 & --37.40 & 0.33 & 0.027 &   5&.8 &B \\
G343.128$-$0.063 &\nodata      & 3.1 & 1665 &  86.65 & --31.74 & 0.38 &  13.64 & --30.69 & 0.33 & 0.024 &  1.8&\rarr 5.5 &C \\
G344.581$-$0.022 &\nodata      & 0.6 & 1665 &   0.82 &    0.90 & 0.60 &   0.57 &  --0.78 & 0.80 & 0.046 &  --2&.8 &B\\*
                 &             &     &      &   2.23 &  --4.40 & 0.69 &  18.57 &  --2.33 & 0.58 & 0.060 &  0.7&\rarr 3.9 &C \\*
                 &             &     &      &  14.33 &  --2.62 & 0.68 &   0.85 &    1.34 & 0.34 & 0.110 &  4.2&\rarr 6.7 &C \\
G345.003$-$0.224 &\nodata      & 2.9 & 1720 &  52.64 & --29.24 & 0.33 &   2.16 & --28.80 & 0.44 & 0.008 &   1&.9 &B \\
G345.011$+$1.792 &\nodata      & 2.2 & 1665 &   0.68 & --18.99 & 0.27 &  20.22 & --20.44 & 0.41 & 0.083 &  --2&.5 &B \\*
                 &             &     &      &  30.84 & --22.74 & 0.37 &  10.32 & --19.72 & 0.25 & 0.105 &  3.3&\rarr 5.1 &C \\
G345.505$+$0.347 &\nodata      & 2.1 & 1665 &  10.91 & --17.32 & 0.35 &   4.37 & --18.01 & 0.27 & 0.106 & --1.2&\rarr --1.8 &C \\*
                 &             &     & 1667 &   2.37 & --12.74 & 0.35 &   4.98 & --12.69 & 0.32 & 0.013 &   0&.1 &B \\*
                 &             &     &      &   1.15 & --21.58 & 0.37 &   3.78 & --21.25 & 0.31 & 0.011 &   0&.9 &B \\
G345.699$-$0.090 &\nodata      & 1.6 & 1665 &   0.52 &  --8.47 & 0.29 &   3.58 &  --8.22 & 0.22 & 0.091 &   0&.4 &B \\
G347.628$+$0.149 &\nodata      & 9.8 & 1612 &   3.02 & --97.16 & 0.27 &   5.89 & --96.42 & 0.27 & 0.014 &   3&.1 &B \\
G348.549$-$0.978 &\nodata      & 2.2 & 1665 &   2.72 & --19.87 & 0.46 &   6.52 & --19.84 & 0.44 & 0.040 &  0.1&\rarr 11.5 &C \\*
                 &             &     & 1720 &   4.74 & --12.77 & 0.36 &   4.49 & --13.38 & 0.38 & 0.036 &  --2&.6 &B \\
G350.011$-$1.341 &\nodata      & 3.1 & 1665 &   5.33 & --19.74 & 0.29 &   3.56 & --19.32 & 0.24 & 0.023 &   0&.7 &B \\
G351.161$+$0.697 & NGC 6334 B  & 2.3 & 1667 &  20.11 &  --9.72 & 0.32 &  78.79 &  --9.64 & 0.22 & 0.028 &   0&.2 &B \\*
                 &             &     &      &   7.37 & --15.29 & 0.29 &   5.08 & --15.24 & 0.30 & 0.006 &   0&.1 &B \\
G351.416$+$0.646 & NGC 6334 F  & 2.0 & 1665 & 184.98 &  --8.87 & 0.32 &  31.40 & --11.99 & 0.57 & 0.045 &  --5&.3 &B \\*
                 &             &     & 1667 &  47.77 &  --9.26 & 0.27 &  54.05 & --11.11 & 0.28 & 0.022 & --2.2&\rarr --5.2 &C\\*
                 &             &     & 1720 &  84.48 &  --9.84 & 0.32 &  61.20 & --10.58 & 0.34 & 0.020 &  --3&.1 &B \\
G351.582$-$0.352 &\nodata      & 6.7 & 1665 &   2.73 & --90.97 & 0.32 &   5.82 & --93.87 & 0.25 & 0.019 &  --4&.9 &B \\
G351.775$-$0.538 &\nodata      & 2.7 & 1665 &   2.94 & --25.62 & 0.46 &   4.80 & --27.86 & 0.39 & 0.025 &  --3&.8 &A \\*
                 &             &     &      & 777.34 &  --1.95 & 0.40 & 106.28 &  --1.85 & 0.47 & 0.013 &   0&.2 &B \\*
                 &             &     &      &   4.09 &  --8.04 & 0.29 &   6.92 &  --7.84 & 1.06 & 0.063 &   0&.3 &B \\*
                 &             &     &      &  54.35 &  --6.88 & 0.40 &   4.42 & --10.16 & 0.57 & 0.047 & --5.6&\rarr --6.5 &C\\*
                 &             &     & 1667 &   3.06 & --26.15 & 0.43 &   4.23 & --27.44 & 0.44 & 0.014 &  --3&.7 &A \\*
                 &             &     &      &  58.40 &  --6.97 & 0.30 &   1.85 &  --4.95 & 0.38 & 0.029 &   5&.7 &B \\*
                 &             &     &      &  10.02 &  --5.55 & 0.27 &   2.62 &  --5.83 & 0.29 & 0.024 &  --0&.8 &B \\*
                 &             &     &      &   1.51 &    0.37 & 0.42 &   1.49 &  --0.16 & 0.67 & 0.037 & --0.3&\rarr --1.5 &C\\
G353.410$-$0.361 &\nodata      & 3.8 & 1665 &  19.83 & --19.56 & 0.29 &   3.30 & --19.50 & 0.48 & 0.033 &  0.1&\rarr  2.4 &C\\
G355.345$+$0.146 &\nodata      &23.1 & 1665 &  17.07 &   19.71 & 0.42 &  15.34 &   16.54 & 0.86 & 0.006 & --3.4&\rarr  --5.4 &C\\
G359.138$+$0.032 &\nodata      & 3.1 & 1665 &   4.60 &  --0.21 & 0.40 &   3.32 &  --2.98 & 0.71 & 0.014 &  --4&.7 &B \\*
                 &             &     &      &   2.92 &  --6.20 & 0.37 &   1.39 &  --6.09 & 0.31 & 0.047 &   0&.2 &B \\
G359.436$-$0.103 &\nodata      & 8.2 & 1665 &   4.66 & --51.83 & 0.48 &   2.60 & --52.11 & 0.63 & 0.032 &  --0&.5 &B  
\enddata
\tnt{a}{Kinematic distances assuming $R_0 = 8.5$ kpc and $\Theta_0 =
220$\kms, unless note in text (\S \ref{sourcenotes}) says otherwise.}
\tnt{b}{Peak flux density, center velocity, and line width (FWHM) of left- and right-circular polarized lines, respectively, 
        as determined by Gaussian fit.}
\tnt{c}{Angular offset between LCP and RCP lines.  When a range of values is given for the magnetic field ($B$), the offset is stated
        for the lines of peak flux density in each polarization.}
\tnt{d}{Full magnitude of $B$ field.  The sign indicates the direction of the line-of-sight component (positive pointing away
        from Sun).  The ratio of the velocity separation between RCP and LCP components to $B$ field was taken to be 0.590, 0.354,
        0.236, and 0.236 \kms $\,$mG$^{-1}$ for the 1665, 1667, 1612,
and 1720 MHz transitions, respectively.  See \S \ref{splitting}
for further details.  When a range is given,
        the extrema represent the minimum and maximum field strength implied by all possible pairings of RCP and LCP lines.}
\tnt{e}{See text (\S \ref{criteria}) for description.}
\tnt{f}{Likely spurious.  See \S \ref{sourcenotes} for details.}

\end{deluxetable}

\begin{deluxetable}{llrcrrcrrcc}
\tabletypesize{\scriptsize}

\tablewidth{0pt}
\tablecaption{Pairs with no Splitting\label{zero_tab}}
\tablehead{
  &&&&\multicolumn{3}{c}{\myhrulefill LCP \myhrulefill}&\multicolumn{3}{c}{\myhrulefill RCP \myhrulefill}&\\
  \colhead{\phm{Source}} &  \colhead{\phm{Alias}} &   
  \colhead{Dist.\tnm{a}} & \colhead{Freq.} &
  \colhead{$S$\tnm{b}} &  \colhead{$v\lsr$\tnm{b}} & \colhead{$\Delta v$\tnm{b}} &
  \colhead{$S$\tnm{b}} &  \colhead{$v\lsr$\tnm{b}} & \colhead{$\Delta v$\tnm{b}} &
  \colhead{$\Delta\theta$\tnm{c}} \\
  \colhead{Source} &   \colhead{Alias} &   \colhead{(kpc)} &
  \colhead{(MHz)} &
  \colhead{(Jy)} &     \colhead{(\kms)} &  \colhead{(\kms)} &
  \colhead{(Jy)} &     \colhead{(\kms)} &  \colhead{(\kms)} &
  \colhead{(arcsec)} 
}
\startdata

G002.143$+$0.010   & \nodata  & 7.5 & 1667 &   2.09 &  59.27 & 0.34 &  0.96 & 59.27  & 0.50 & 0.026 \\
G009.622$+$0.195   & \nodata  & 5.7 & 1667 &   4.67 &   7.13 & 0.29 &  0.92 &  7.14  & 0.36 & 0.032 \\
G043.167$+$0.010   & W49 N    &11.5 & 1665 &  10.05 &   7.88 & 0.36 & 32.08 &  7.88  & 0.43 & 0.014 \\
G080.864$+$0.421   & \nodata  & 4.1 & 1665 &  20.10 & --8.46 & 0.26 & 26.02 & --8.46 & 0.24 & 0.009 \\
G081.721$+$0.571   & W75 S    & 2.0 & 1667 &   0.74 & --1.12 & 0.27 &  1.56 & --1.14 & 0.37 & 0.070 \\
G081.871$+$0.781   & W75 N    & 2.0 & 1665 &  26.00 &   3.09 & 0.35 & 29.77 &  3.07  & 0.36 & 0.002 \\
                   &          &     &      &  13.95 &   0.63 & 0.39 & 17.31 &  0.64  & 0.39 & 0.003 \\
G111.533$+$0.757   & NGC 7538 & 2.8 & 1665 &   0.94 &--57.69 & 0.30 &  0.64 &--57.70 & 0.35 & 0.011 \\
G133.946$+$1.064   & W3 OH    & 2.2 & 1665 &  32.94 &--47.45 & 0.22 & 20.61 &--47.45 & 0.22 & 0.015 \\
G350.011$-$1.341   & \nodata  & 3.1 & 1667 &   2.25 &--23.74 & 0.51 &  2.60 &--23.78 & 0.50 & 0.016 \\
G351.775$-$0.538   & \nodata  & 2.7 & 1665 &  40.53 & --9.24 & 0.25 & 37.35 & --9.23 & 0.27 & 0.027 \\
                   &          &     &      &  11.66 &   1.22 & 0.50 & 11.66 &  1.22  & 0.43 & 0.099 
\enddata
\tnt{a}{Kinematic distances assuming $R_0 = 8.5$ kpc and $\Theta_0 = 220\,$ \kms, unless note in text says otherwise.}
\tnt{b}{Peak flux density, center velocity, and line width (FHWM) of left- and right-circular polarized lines,
        respectively, as determined by Gaussian fit.}
\tnt{c}{Angular offset between LCP and RCP lines.}

\end{deluxetable}

\section{Zeeman Pairs}

\subsection{Distances\label{distances}}

The distances given in Tables 1 and 2 are kinematic distances assuming
a flat rotation curve with $R_0 = 8.5$~kpc and $\Theta(R) = 220$~\kms,
except as noted in the text.  For sources inside the solar circle in
the first and fourth Galactic quadrants, there is a distance
ambiguity.  For a number of these sources, the kinematic distance
ambiguity is resolved from \ion{H}{1} absorption observations by
\citet{fish}.  That paper also provides an analysis of the scale
height of known UC\ion{H}{2} regions (embedded in massive SFRs) above
the Galactic plane, including an analysis of the accuracy of resolving
the distance ambiguity to a UC\ion{H}{2} region given only the
kinematic distances and Galactic \emph{latitude} of the region.  When
the distance was not known by any other method, the Galactic latitude
was used to resolve the kinematic distance ambiguity toward
massive SFRs only when the difference between the near and far
kinematic distances exceeds 4.5 kpc, corresponding to 90\% predictive
accuracy \citep{fish}.  The majority of these sources are located in
the fourth Galactic quadrant.

When evaluating kinematic distances, we assumed that massive SFRs have
a peculiar motion of $\lesssim 10$~\kms.  This freedom is important to
placing a few sources (e.g., G355.345$+$0.146) at a sensible distance.
Nevertheless, it is possible for peculiar motions to exceed 10 \kms.
Hydroxyl and H$_2$CO absorption measurements toward G10.624$-$0.385
and other sources in the W31 complex demonstrate deviations from the
velocities expected from the rotation model of as much as 36 \kms\
\citep{wilson}.  

For sources whose median LSR velocity of OH maser emission is close to
0 \kms, the near kinematic distance is often unrealistically close.
Since the angular window radius as described in \S \ref{criteria}
is inversely proportional to the assumed distance to the source, a
source assumed to be much nearer than it is would be partitioned into
windows encompassing too large of an angular area.  The resulting
windows would be few in number and contain a large number of maser
lines.  The effect is amplified for especially near distances (a few
hundred parsecs or less), where a small velocity error translates into
a large fractional distance error and therefore a large fractional
error in the window size.  To be consistent (if somewhat arbitrary) in
handling these extremely near sources ($\ll 2$~kpc), we used the
distance produced by averaging the kinematic distance derived from the
median OH maser velocity and 2.0 kpc.  Since massive SFRs are few and
far between, it is unlikely that a massive SFR whose distance is not
independently known is actually located within 1~kpc of the Sun.

\subsection{Satellite-Line Field Splitting\label{splitting}}

Unlike the main-line (1665 and 1667 MHz) transitions, which split into
only one pair of $\sigma$-components, the satellite-line transtitions
split into three pairs with line intensities in the ratio 6:3:1
\citep{davies}.  As mentioned in a footnote to Table \ref{pairs_tab},
the Zeeman splitting coefficient for the satellite-line (1612 and 1720
MHz) transitions was assumed to be 0.236 \kms$\,$ mG$^{-1}$.  This
splitting coefficient is correct for the intensity-weighted average of
the three lines in each polarization.  However, maser amplification
may be greatest for the strongest line in each triad.  If only the
strongest line is seen after amplification, the splitting coefficient
would be 0.114 \kms$\,$ mG$^{-1}$, or approximately half of the
intensity-weighted value.  Indeed, the splitting coefficient could be
between the two extremal values, as would be consistent with
amplification of all three $\sigma$-components in a
partially-saturated maser.  A comparison of the magnetic field
strengths inferred from 1720 MHz pairs in W3(OH) listed in Table
\ref{pairs_tab} with those shown on VLBI maps \citep{w3oh} is more
consistent with a splitting coefficient closer to 0.236 \kms$\,$
mG$^{-1}$.  Analysis of field strengths inferred from 1612 and
1720~MHz Zeeman pairs in all sources also supports a large splitting
coefficient (see \S \ref{strengths}).  In any case, the direction
of the line-of-sight component of the field is unaffected.

\subsection{Notes on Specific Sources\label{sourcenotes}}

G000.670$-$0.058:  The separation between the LCP and RCP components of
the pair exceeds the assumed $3 \times 10^{15}$~cm correlation length
by less than a factor of two.

G005.886$-$0.393:  The two pairs receiving an A grade are separated by
approximately 0\arcsec.3.

G009.622$+$0.195: We adopt a distance of 5.7~kpc to this source,
consistent with observations by \citet{scoville} and \citet{hofner},
although the distance obtained kinematically could be as near as
0.7~kpc.  The second Zeeman pair listed in Table \ref{pairs_tab},
offset by about $5\arcsec$ from the other two pairs, is in a separate
maser site within the same massive star-forming cluster.  Additional
lines for the first pair in table 1 receiving a grade of C consist of
an LCP line of 1.65 Jy at $-3$.97 \kms\ and an RCP line at 7.85 Jy and
1.22 \kms.  For the second such pair, there are four additional LCP
lines: 0.79 Jy at 7.04 \kms, 0.39 Jy at 6.50 \kms, 0.47 Jy at 6.09
\kms, and 1.32 Jy at 5.49 \kms.

G012.680$-$0.181: Additional lines for the 1665 pair receiving a grade
of C consist of two RCP lines: 2.23 Jy at 60.71 \kms\ and 2.58 Jy at
61.17 \kms.  For the 1667 pair, there are two additional LCP lines of
2.92 Jy at 64.81 \kms\ and 2.28 Jy at 62.76 \kms\ and one additional
RCP line of 3.30 Jy at 62.15 \kms.  From observations using the 43 m
telescope at Green Bank, \citet{zhengw33} reports two Zeeman pairs of
around $-5$ mG each.  The 4 kpc distance assumed for W33 B and W33 A
is the kinematic distance taken from \citet{haschick}.

G020.081$-$0.135:  There is an additional LCP line of 4.44 Jy at 45.47
\kms\ for the pair receiving a C grade.

G032.744$-$0.076:  There is an additional RCP line of 0.24 Jy at 40.24
\kms\ for the pair receiving a C grade.

G034.257$+$0.154:  There is an additional LCP line of 2.24 Jy at 57.97
\kms\ for the pair receiving a C grade.  This source is one of three
in the G34.3$+$0.2 complex.  The magnetic field structure of this
cometary \ion{H}{2} region, source C in \citet{zrm}, is unclear.  The
magnetic field of source B (G034.258$+$0.153) is ordered and pointing
primarily away from the Sun.

G035.024$+$0.350:  There is an additional LCP line of 0.54 Jy at 43.83
\kms\ and an RCP line at 1.83 Jy and 44.63 \kms.

G035.197$-$0.743:  There is an additional LCP line of 1.04 Jy at 26.35
\kms.

G035.577$-$0.029:  In each of the two pairs, the angular offset between
the LCP and RCP lines is consistent with a separation of about $2
\times 10^{16}$~cm, assuming the kinematic distance.  There are two
additional LCP lines of 2.85 Jy at 51.89 \kms\ and 2.61 Jy at 49.91
\kms\ for the second pair receiving a D grade.

G040.622$-$0.137:  There is an additional RCP line of 1.04 Jy at 34.76
\kms\ for the 1665 pair receiving a grade of C.  For the 1667 pair,
there are two additional LCP lines: 2.82 Jy at 31.53 \kms\ and 1.34 Jy
at 30.85 \kms.

% G043.148$+$0.015:  [W49 -- circle size]

G045.071$+$0.134:  For the 1665 pair, there are two additional LCP
lines of 6.24 Jy at 53.57 \kms\ and 8.53 Jy at 52.95 \kms\ and one
additional RCP line of 7.21 Jy at 55.91 \kms.  For the first 1667
pair, there is an additional RCP line of 7.21 Jy at 55.91 \kms.  For
the second 1667 pair, there is an additional LCP line of 2.97 Jy at
55.78 \kms\ and an RCP line of 0.91 Jy at 57.17 \kms.

G045.465$+$0.047:  The LCP and RCP lines of each of the two pairs are
located in a circle of $3 \times 10^{15}$~cm.  While the pairing of
these lines is theoretically ambiguous, we feel that the identical
magnetic fields implied by the listed pairings and the small velocity
difference between these pairings (0.65 \kms, less than the 0.8 \kms\
turbulent velocity differences expected within a masing cloud) are
sufficient to justify our claimed magnetic field values.

G049.469$-$0.370:  In 18-cm VLA observations, \citet{gaume} also find a
Zeeman pair with a positive magnetic field.  For more information, see
the notes for the next source.  

G049.488$-$0.387:  \citet{gaume} find about half a dozen
Zeeman pairs in the W51 complex around regions e1 and e2, all
indicating a positive magnetic field.  In 5-cm VLBI observations,
\citet{desmurs} find a Zeeman pair between regions e1 and e2.
However, using the VLBA \citet{arm2} find two very strong 1665 MHz
Zeeman pairs indicating magnetic fields of $-21$ and $-19.8$~mG.  The
kinematic distance is taken from \citet{caswell}. For the 1665 pair,
there is an additional LCP line of 59.42 Jy at 58.90 \kms.

G069.540$-$0.976:  For the 1665 pair receiving a C grade, there are two
additional LCP lines: 5.07 Jy at 15.45 \kms\ and 3.17 Jy at 15.06
\kms.  Preliminary (as yet unpublished) VLBA 18-cm observations by the
authors of this paper and VLBI 5-cm maps of \citet{desmurs} confirm
that the magnetic field is consistently oriented toward the Sun across
the source.

G070.293$+$1.601:  For the 1665 pair receiving a C grade, there is an
additional LCP line of 4.03 Jy at $-20$.39 \kms.  For the 1667 pair,
there is an additional LCP line of 2.46 Jy at $-20$.80 \kms.  However,
the two grade-C pairs in the table meet the self-consistency criteria
($\Delta \theta < 0\arcsec\!\!.2$, $\Delta v_c < 0.2$
\kms, $\Delta B <$ 0.4 mG between pairs in the two transitions).

G081.721$+$0.571:  Both this source and G081.871$+$0.781, believed to
be in the same complex, have average LSR velocities of a few \kms\
positive.  This puts these sources near the tangent point at this
Galactic longitude with a large uncertainty in distance.  We adopt the
\citet{dickel} distance of 2.0 kpc as the distance to the W75 complex.
For the pair receiving a grade of C, there are additional LCP lines of
15.48 Jy at 0.93 \kms, 8.45 Jy at 0.58 \kms, and 2.44 Jy at $-0$.27
\kms\ as well as RCP lines of 0.40 Jy at 4.79 \kms, 0.65 Jy at 3.53
\kms, and 1.49 Jy at 2.97 \kms.

G081.871$+$0.781:  Preliminary (as yet unpublished) VLBA observations
indicate a reversal of the line-of-sight magnetic field direction
across the source.  The $-25.4$~mG pair is not seen with the VLBA.
However, a cluster that contains a large number of maser spots with a
large velocity spread ($\approx 20$~\kms) is detected.  It is probable
that this ``pair'' is actually two nearby lines whose velocity
difference is due to a velocity gradient rather than an extremely
large magnetic field.  For distance information, see the note for
G081.721$+$0.571.

G109.871$+$2.114:  For the 1665 pair receiving a C grade, there is an
additional LCP line of 13.65 Jy at $-13.$90 \kms.  However, the
ambiguity is resolved by examining the 1667 transition, which is
within 0\arcsec.2 of the 1665 pair in question.  \citet{cohen} find a
Zeeman pair that implies a magnetic field direction oriented away from
the Sun.  Monitoring of this Zeeman pair over time shows that the
strength of the magnetic field is decreasing.

G111.533$+$0.757:  The distance of 2.8 kpc is taken from
\citet{campbell}.

G133.946$+$1.064:  For the 1612 pair, there is an addition RCP line of
3.76 Jy at $-42$.92 \kms.  For the 1720 line receiving a C grade,
there is an additional LCP line of 6.88 Jy at $-45$.23 \kms\ and two
additional RCP lines: 4.47 Jy at $-42$.72 \kms\ and 7.47 Jy at
$-44$.36 \kms.  Previous VLBI observations have clearly established
that the magnetic field is predominantly oriented away from the Sun
\citep{garcia}.  The 2.2 kpc distance used for W3(OH) is based on OB
star luminosities from \citet{humphreys}.

G343.128$-$0.063:  There is an additional LCP line of 54.52 Jy at
$-33$.93 \kms.

G344.581$-$0.022:  The near kinematic distance of 0.6~kpc
\citep{forster} is too close, as evidenced by the cluttered spectra
produced in circles of radius $3\times 10^{15}$~cm, but no pairs are
seen at all at the far kinematic distance of 15.8~kpc.  A distance of
2.0 kpc was used, which produces circles of radius $10^{15}$~cm
assuming the near kinematic distance is correct.  For the first pair
receiving a C grade, there is an additional LCP line of 1.99 Jy at
$-4$.66 \kms\ and an RCP line of 2.93 Jy at $-4$.01 \kms.  For the
second such pair, there is an additional LCP line of 3.74 Jy at
$-1$.16 \kms.

G345.011$+$1.792:  For the pair receiving a C grade, there is an
additional LCP line of 3.39 Jy at $-21$.67 \kms.

G345.505$+$0.347:  For the pair receiving a C grade, there is an
additional LCP line of 2.72 Jy at $-16$.92 \kms.

G348.549$-$0.978:  For the pair receiving a C grade, there are two
additional RCP lines: 4.46 Jy at $-13$.11 \kms\ and 2.64 Jy at
$-19$.26 \kms.

G351.416$+$0.646:  For the pair receiving a C grade, there is an
additional LCP line of 22.29 Jy at $-10$.33 \kms.

G351.775$-$0.538:  For the 1665 pair receiving a C grade, there is an
additional RCP line of 3.04 Jy at $-10$.72 \kms.  For the 1667 pair
receiving a C grade, there is an additional RCP line of 1.11 Jy at
0.25 \kms.  Preliminary VLBA observations indicate a reversal of the
line-of-sight magnetic field direction across the source \citep{arm2}.

G353.410$-$0.361:  There is an additional LCP line of 3.55 Jy at
$-20$.56 \kms\ and an RCP line of 2.34 Jy at $-19$.12 \kms.
\citet{caswell} finds evidence for a magnetic field pointing in the
opposite direction based on 6030- and 6035-MHz OH masers.

G355.345$+$0.146:  The rotation curve is quite flat in this part of the
galaxy.  The distance of 23.1~kpc was derived assuming a velocity of
$+7.5$ \kms, or 10 \kms\ less than the actual average value as deduced
from OH maser emission.  (Recall that a positive velocity in the
fourth Galactic quadrant indicates that the source is located outside
the solar circle, or at least 17 kpc away at this Galactic longitude.)
The distance deduced from the actual average velocity of $+17.5$ \kms\
is unphysically large, assuming a flat rotation curve with $R_0 =
8.5$~kpc and $\Theta(R) = 220$~\kms.  There are two additional LCP
lines: 15.38 Jy at 18.95 \kms\ and 15.84 Jy at 18.55 \kms.

\subsection{Distribution of Magnetic Field Strengths\label{strengths}}

A histogram of the inferred magnetic field strengths for 75 Zeeman
pairs from Table \ref{pairs_tab} is included in Figure
\ref{histogram}.  Only those Zeeman pairs implying a single value of
the magnetic field strength are included, and the $-25.4$~mG pair in
G081.871+0.781 is omitted (see \S \ref{sourcenotes}).  A Gaussian fit
is superimposed on the histogram.  The distribution of field strengths
has a HWHM of $3.9 \pm 0.1$~mG and is centered at 0~mG to within the
formal error.  This is consistent with average field magnitudes of
several milligauss deduced from OH Zeeman splitting in other studies
\citep[e.g.,][]{garcia}.  Note that this field strength is what would
be expected for magnetic field strength enhancement of $|B| \propto
n^{1/2}$ \citep{mouschovias} from interstellar values of $B$ and $n$
($\sim 1~\mu$G, $\sim 1$~cm$^{-3}$) to $n \sim 10^6$~cm$^{-3}$, as is
typical for sites of OH masing \citep{reidmoran}.

Of the 15 Zeeman pairs detected at 1612 and 1720~MHz, 10 imply a
magnetic field strength less than or equal to the HWHM assuming a
splitting coefficient of 0.236 \kms$\,$ mG$^{-1}$, 3 imply a field
strength greater than the HWHM, and the 2 pairs with a grade of C in
Table \ref{pairs_tab} are inconclusive owing to pairing ambiguities.
If a splitting coefficient of 0.114 \kms$\,$ mG$^{-1}$ is assumed, 3
Zeeman pairs imply a magnetic field strength less than the HWHM and 12
imply a greater field strength.  Thus, magnetic field strengths
deduced from 1612 and 1720~MHz Zeeman pairs are more consistent with
values obtained from the main-line transitions when a splitting
coefficient value closer to the high value (0.236 \kms$\,$ mG$^{-1}$)
is assumed.  While we cannot rule out the possibility that the
physical requirements for masing at 1612 and 1720~MHz cause
satellite-line maser spots to be observed preferentially at locations
of increased magnetic field strength, we feel justified in using the
larger splitting coefficient.

%HWHM: 3.85 +/- 0.06
%  mu: -0.05 +/- 0.05
% Excluding 2 in -1<->0 bin and 6 in 0<->1 bin:
%HWHM: 3.80 +/- 0.09
%  mu: -0.11 +/- 0.08

\subsection{Other OH Zeeman Studies\label{otherzeeman}}

This study was undertaken primarily to investigate Davies's hypothesis
that the magnetic fields deduced from OH Zeeman splitting in
massive SFRs exhibit Galactic-scale organization.
Preliminary analysis of the data from the Paper I survey yielded 40
massive SFRs for which OH Zeeman splitting suggested a local
field direction \citep{fram}.  While the data set was homogeneous, it
excluded southern sources, which are not observable from the VLA.  To
address this, the current study includes OH Zeeman pairs from a number
of other studies as well.  Chief among these are the 6 GHz surveys of
\citet{baudry} and \citet{caswellv}, the 1.6 GHz study of
\citet{gaume}, and the original literature search of
\citet{reidsilver}.  Two sources for which we have unpublished maps in
the 1665 and 1667~MHz OH transitions with the VLBA have also been
included (ON2~N and S269) (in preparation).  Other supplementary
papers relevant to only one or a small number of our sources are
mentioned with the appropriate source(s) in \S \ref{sourcenotes}.
The objective is to compile a list of massive SFRs restricted to those
with a predominant magnetic field direction.

%Preliminary VLBI observations of some UC\ion{H}{2} regions show that
%organized line-of-sight magnetic field reversals exist across some
%individual sources \citep{arm2}.  Nevertheless, the field structure
%around UC\ion{H}{2} regions appears to be ordered, showing either a
%consistent line-of-sight field direction across the entire source or a
%single reversal.  These ordered reversals may be explained by analogy
%to a magnetic field around a conducting region.  The UC\ion{H}{2} acts
%as a conductor, and any ambient large-scale magnetic field in the
%neutral material must bend around the ionized region.  In the simplest
%case of a conducting sphere immersed in a large-scale uniform magnetic
%field, the field lines bend along the surface of the conducting region
%as shown in Figure ***** \citep{bourke}.  Depending on the angle
%between the magnetic field direction and the vector toward the
%observer, the line-of-sight field projection will appear to reverse
%across the source.

When assigning an overall magnetic field direction to a massive SFR,
we require that at least two-thirds of all Zeeman pairs for any one
source imply the same line-of-sight field direction.  In addition to
problems associated with a possible magnetic field reversal across the
source, the VLA beam is large enough to encompass several maser
clumps.  While this could potentially blend maser spots, resulting in
misidentification of Zeeman pairs, we have striven to obtain a sound
collection of Zeeman pairs by discarding all pairs that do not meet
the objective criteria outlined in \S \ref{criteria}, at the
expense of completeness.  The even lower spatial resolution of the
data taken from $\lambda = 5$~cm studies is not a concern since the
intensity of RCP and LCP lines in a 6035~MHz pair is typically much
more equal than is seen in ground-state Zeeman pairs \citep{caswellv}.
This fact, combined with a much smaller Zeeman splitting coefficient,
makes unambiguous pairing of maser spots easier in the 6~GHz
transitions.  Thus we do not believe that blending of multiple maser
spots in the same beamwidth is a significant source of contamination
for the magnetic field measurements in this study.  The final sample
to be used in \S \ref{sgmf} comprises 45 of the 53 sources in
Table \ref{pairs_tab}, plus 29 from other information.

\section{Discussion}

A number of historical findings motivated us to study whether or not
the Galactic magnetic field is sensed by OH maser Zeeman splitting.
The original hypothesis of a link between the Galactic magnetic field
and interstellar OH masers was advanced by \citet{davies}, who, noting
that his compilation of eight OH maser sources all displayed a magnetic
field consistent with the clockwise direction of Galactic rotation,
suggested that the OH masers traced a circular Galactic magnetic
field.  For the most part, the OH sources included in his study are
located near the solar circle, precluding investigation of reversals
of the large-scale magnetic field direction with Galactocentric
radius.  \citet{reidsilver} tested the Davies hypothesis by finding 12
\emph{new} OH maser sources, of which 10 had a line-of-sight magnetic
field aligned with the direction of Galactic rotation.  This seemed to
be consistent with Davies's suggestion and a concentric-ring model of
the Galactic magnetic field as deduced from the pulsar rotation
measure study of \citet{randk}.  However, Rand \& Kulkarni also
presented evidence for magnetic field reversals between rings, which
was not critically tested in the Reid \& Silverstein study.

\subsection{Preservation of the Galactic Magnetic Field Direction}

Should the magnetic field as deduced from the Zeeman splitting of OH
masers around massive SFRs show any Galactic-scale structure?
This would require three important conditions to be satisfied.  First,
there must be a large-scale Galactic magnetic field, which is likely
given pulsar \citep[e.g.,][]{manchester} and extragalactic rotation
measure observations \citep[e.g.,][]{sofue}.  Second, the OH clumps
must ``remember'' the Galactic magnetic field after a magnetic
compression of about three orders of magnitude from interstellar
values.  And third, intra-source reversals of the magnetic field
in an individual massive SFR must be sufficiently few so as
to not destroy a possible observed large-scale Galactic magnetic field
by the inclusion of inferred local magnetic field directions that are
opposite to the actual prevailing magnetic field direction near the
source.

Of these three conditions, the most difficult to justify on a physical
basis is the assumption that the magnetic field direction at a site of
massive star-formation would retain its initial (Galactic) direction
through collapse.  Ammonia observations of W3(OH) suggest that the
density in OH maser clumps is $n \approx 5 \times 10^6$~cm$^{-3}$
\citep{rmb}, which is a factor of $10^{3-4}$ higher than found in
giant molecular clouds and a factor of $10^6$ greater than typical
values in the ISM.  It is unclear why the magnetic field in these
clumps would retain the orientation of the field in the giant
molecular cloud (GMC) during the compression phase, which may also
entail many rotations of the clumps around the central condensation,
and why the GMC would retain the orientation of the interstellar
field.

\subsection{Structure of the Galactic Magnetic Field\label{sgmf}}

Assuming that the second and third of the above conditions are
satisfied, we can investigate whether there exists a large-scale
Galactic magnetic field and what form it takes.  The simplest
imaginable Galactic field would be a circular field with no reversals
in the portion of the disk where massive SFRs are found.  Our
74-source sample discussed in \S \ref{otherzeeman} do not support
this model.  A plot of the predominant sense of the magnetic field
direction (i.e., clockwise or counterclockwise as viewed from the
North Galactic Pole) deduced from the OH masers near massive SFRs in
this study is shown in Figure \ref{galaxy}.  For a circular magnetic
field configuration in the Milky Way, the sense of the line-of-sight
orientation (directed toward or away from the Sun) flips across the $l
= 0, l = 180\degr$ line.  Noting this, 41 have magnetic fields
directed in a clockwise sense as viewed from the North Galactic Pole
and 33 have fields directed in a counterclockwise sense.  This does
not support a uniform circular field model without reversals.  Indeed,
it would not be inconsistent with a completely random distribution of
magnetic field orientations.

The rough equality of sources with clockwise and counterclockwise
fields does not preclude an organized field structure.  For instance,
the magnetic field could be aligned along either concentric rings or
spiral arms but with reversals between the rings or arms.
Alternatively, the magnetic field as traced by interstellar OH masers
could exhibit organized structure in small (kiloparsec-scale) patches,
but not on a Galactic scale.  Since our data do not imply a uniform
sense of the Galactic magnetic field, a two-point spatial correlation
function was calculated on the data in order to test the hypothesis
that magnetic field directions of massive SFRs exhibit some
degree of correlation over kiloparsec scales.  We define the magnetic
field orientation of two massive SFRs to be correlated if they
are both oriented in a clockwise or counterclockwise Galactic sense.
We similarly define a pair to be anti-correlated if the pair consists
of one source with a clockwise field and one with a counterclockwise
field.  The correlation function was defined as the number of pairs of
massive SFRs whose fields were correlated in the above sense,
divided by the total number of pairs of sources.  This function was
calculated for a range of intersource distances.  Since interpretation
of the significance of the correlation function is difficult in
isolation, it was compared against the same function evaluated for
randomly-generated magnetic-field direction data at the locations of
the massive SFRs in our sample.  In the random sample, the
positions of the maser sources were held fixed, but the line-of-sight
field orientation of each source was randomly assigned as toward or
away from the Sun.  An ensemble of $10^5$ such random field
distributions was produced.  The results are shown in Table
\ref{corr_tab}.  The percentages of trial runs showing a lesser and
greater degree of correlation do not sum to 100\% because some random
distributions of magnetic fields produced identical correlation
results.

Since the data exhibit roughly equal likelihood of less and more
correlation at the 10 or 20~kpc level, our data do not support the
original hypothesis of \citet{davies} that there exists a uniform,
clockwise Galactic field.  On Galactic scales, pulsar rotation measure
studies suggest that field reversals exist between spiral arms
\citep[e.g.,][]{han,rand}.  This would destroy any large-scale
correlation on the Galactic scale, as spiral arms with
oppositely-oriented magnetic fields would be included at this scale.

On sub-kiloparsec scales there are a number of source complexes in
which two nearby massive SFRs have oppositely-oriented
line-of-sight magnetic fields.  The W49 complex provides a good
example of this: G43.148+0.015 \citep[the Zeeman pair is between the
sources W49A/R$_2$ and R$_3$ in the classification of][]{depree} and
W49S have fields directed primarily away from the Sun, while the field
of W49N is directed toward the Sun.  It is possible that this may
result from a tangled field geometry associated with multiple cores
in the same complex.

Nevertheless, we do note a possible field organization on a $\sim
2$-kpc scale.  Indeed, several patches of field coherence at this
scale are noticeable in Figure \ref{galaxy}.  For instance, there is a
cluster of sources with clockwise magnetic fields (i.e., oriented
toward the Sun) near $(X,Y) = (-4,-3)$~kpc, and another cluster with
clockwise fields near $(3,5)$~kpc.

\subsection{Rotation Measure Comparison\label{rmcomp}}

The structure of the Galactic magnetic field can be studied with
rotation measures of Galactic pulsars and extragalactic sources.  We
find that toward the second and third Galactic quadrants, which
contain only six sources, all fields are oriented in the clockwise
direction, while the nearest six sources in the first and fourth
quadrants are consistent with a counterclockwise field.  This agrees
with rotation measure studies of the outer Galaxy, which also find an
overall clockwise field \citep{lynesmith, browntaylor}.  This also
agrees with a reversal inward claimed in other papers from pulsar
rotation measure estimates \citep{simard79,randk,clegg}.

In order to compare our results with the Galactic magnetic field
deduced from rotation measures, we generated simulated rotation
measure data from the line-of-sight magnetic field direction of our OH
maser data.  Our data consist of measurements of milligauss fields
\emph{in situ} around massive SFRs, while extragalactic rotation
measures are line-of-sight integrals predominantly through the
microgauss fields of the interstellar medium.  We calculated simulated
rotation measures along rays of Galactic longitude within the Galactic
plane using the line-of-sight field directions from OH masers.  The
rotation measure $RM$ of radiation coming from a source at a distance
$D$ is given by
\[
RM = 8.1 \times 10^5~\mathrm{rad~m^{-2}} \int_D^0 n_e \, \mathbf{B \cdot dl},
\]
where $n_e$ is the electron density in cm$^{-3}$, $\mathbf{B}$ is the
magnetic field in gauss, and $\mathbf{dl}$ is measured in parsecs.
Typical values for these variables in the interstellar medium are $n_e
\approx 0.03$~cm$^{-3}$ \citep{lynemt} and $B_\| \approx 2 \times
10^{-6}$~G \citep{manchester,simard80,heiles}.  Given that magnetic
fields implied by Zeeman splitting in massive SFRs exhibit some
correlation on a scale of $\lesssim 2$~kpc (see \S \ref{sgmf}),
we postulated that the magnetic field around a massive SFR indicates
the prevailing Galactic magnetic field direction in a sphere of radius
$r$ centered at the massive SFR.  We adopted $r \sim 0.5$~kpc, which
yields a contribution of about 50~rad~m$^{-2}$ to the rotation measure
of a line of sight passing along the diameter of each sphere.  Thus,
we assigned a rotation measure of $\pm 50$~rad~m$^{-2}$ to each
sphere, depending only on the line-of-sight magnetic field direction
in the massive SFR, and summed the rotation measures of the spheres
intersected by the ray.  For simplicity, each sphere was assumed to
contribute $\pm 50$~rad~m$^{-2}$ to the total rotation measure,
regardless of the path length of the ray within the sphere.  Note that
since the radiation propagates from $D$ to $0$, $\mathbf{dl}$ is
oriented toward the Sun, and $\mathbf{B \cdot dl}$ is negative when
the magnetic field is oriented away from the Sun.  Thus, a positive
magnetic field generates a negative rotation measure.

Our simulated rotation measures are plotted in Figure \ref{rml} for $r
= 0.5$~kpc.  Features of the plot are not highly sensitive to the
value of $r$, but much lower values leave a large volume of the
Galactic plane unsampled by our data, while much higher values result
in a lot of overlap of the spheres.  Superposed atop this plot are
extragalactic rotation measures reproduced directly from Figure 9 of
\citet{randk}, based on data from \citet{simard81}.  Our data agree
reasonably well with the extragalactic rotation measure data.  In the
second and third Galactic quadrants, our data match the sign of the
extragalactic rotation measures.  While the extragalactic rotation
measures are not generally of consistent sign in the first and fourth
quadrants, agreement is found again at $l = 15\degr$, where the
extragalactic rotation measures are decidedly positive.  The simulated
rotation measures we deduce from OH Zeeman splitting appear to fit the
extragalactic data at least as well as, if not better than, the ring
model of \citet{randk}.

\subsection{Spiral Arms}

In many models, one would expect the Galactic field to be organized
along spiral arms.  However, investigating whether the Galactic field
shows organization on a per-arm basis necessitates assignment of
massive SFRs to spiral arms.  Based on positions on both a plot of the
Galaxy and on the longitude-velocity diagram in Figure \ref{lvplot},
most of the sources in our study were assigned to a spiral arm, as
listed in Table \ref{arm_tab}.  The above correlation analysis was
repeated on the subset of sources associated with each spiral arm, and
the results are included in Table \ref{corr_tab}.  For the most part,
no obvious correlations exist except for sources assigned to the Norma
arm, where 8 of 10 sources (and all 7 in the fourth quadrant) have a
magnetic field oriented clockwise with respect to the Galactic center.

As noted above, it is difficult to make strong statements pertaining
to field directions in spiral arms.  The shape and locations of the
spiral arms in the Milky Way are still a matter of debate, and many of
the source distances used in this study are kinematic and therefore
subject to large enough errors that some could be placed in the wrong
arms.  Even assuming that the near/far kinematic distance ambiguity
has been resolved correctly for all sources, a typical distance error
by this method is about 1~kpc.  Furthermore, the OH maser velocity may
differ from the velocity of the massive SFR, which may in turn
deviate quite significantly from uniform circular motion, as discussed
in \S \ref{distances}.  Especially near $l = 0\degr$, where
several spiral arms cross in a longitude-velocity diagram, it is
difficult to reliably assign massive SFRs to spiral arms.

\subsection{Intra-source Field Reversals}

The problem of relating OH Zeeman magnetic field measurements to the
Galactic magnetic field can be inverted.  Assuming that there exists a
large-scale Galactic field, do our results allow us to comment either
on whether the magnetic field threading OH maser clumps are related to
the (pre-collapse) local Galactic field or on whether intra-source
field reversals confuse the issue too much to see conclusive evidence
of large-scale field organization?

These are questions that cannot be answered conclusively with the VLA
alone, due to its large beam size.  For instance, no conclusive Zeeman
pairs are found in the 1665 MHz transition in the VLA data for W3(OH).
In contrast, \citet{w3oh} identified 23 Zeeman pairs in the same
transition using VLBI.  Significant blending of independent maser
features within a single VLA beam often renders identification of
Zeeman pairs ambiguous, as discussed in \S \ref{otherzeeman}.
When Zeeman pairs are identified with VLA resolution, they are often
in short supply.  We were able to find only one Zeeman pair in many of
the massive SFRs shown in Table \ref{pairs_tab}.  One pair does
not conclusively determine the prevailing magnetic field direction
surrounding the region.  Even in the simple case, depicted in Figure
\ref{sphere}, of a conducting, spherical \ion{H}{2} region in a
uniform magnetic field, the line-of-sight projection of the magnetic
field direction may vary depending on the inclination of the source
\citep{bourke}.  A numerical simulation was run to estimate the
fraction of sources whose line-of-sight magnetic field would be
incorrectly inferred owing to projection effects for this simple
geometry.  Given an ensemble of ideal, spherical \ion{H}{2} regions,
each with a randomly-oriented prevailing magnetic field direction and
a single magnetic field measurement taken locally at a random point on
the projection of the sphere (as from a Zeeman pair), the
line-of-sight direction of the measured field will be opposite to the
line-of-sight direction of the prevailing magnetic field 25\% of the
time.

Intra-source field reversals have been seen across three of nine
massive SFRs that were mapped in the main-line OH transitions with the
VLBA: G351.775$-$0.538 \citep{arm2}, W75N, and W75S (unpublished
data).  Thus, one-third of the population of massive SFRs mapped at
milliarcsecond resolution show at least two distinct maser clumps with
Zeeman pairs indicating opposite line-of-sight directions.  While one
might have expected that the magnetic field would be oriented randomly
in a massive SFR owing to a complex evolutionary history that
``scrambles'' the pre-collapse field, massive SFRs with intra-source
field reversals still exhibit an organized field structure.  More
likely, these reversals are a projection effect due to the geometry of
the field lines relative to the observer.  Still, the ionized sphere
model is surely an oversimplification in some cases.  For instance,
the OH masers around W75N are aligned along two intersecting lines,
one running roughly north-south and the other east-west \citep{baart}.
Our VLBA data show that the magnetic fields along the north-south line
point consistently away from the Sun, while those in the east-west
line point consistently toward the Sun.  Baart et al.\ suggest that
the masers trace expanding shock waves, while \citet{slysh} claim to
have found evidence that the north-south structure is a thin, rotating
disk.  Whatever the explanation of this source may be, the observed
intra-source field reversal is clearly not a simple projection effect
in a static environment.

\subsection{The Random Component of the Galactic Magnetic Field}

Even if the Galactic magnetic field consists of a large-scale field
with regular structure, the turbulent field component may hinder clear
detection of the pattern.  Synchrotron polarization observations
suggest that the ratio of field strengths of the regular to turbulent
magnetic fields may be between 0.6 and 1.0, with lower values more
likely in spiral arms \citep{beck}.  Assuming a uniformly random
two-dimensional (in the Galactic plane) orientation of the turbulent
component of the magnetic field, the line-of-sight orientation of the
total field will be opposite to that of the regular component 8\% to
20\% of the time for the aforementioned range of field strength
ratios.  Thus, even if the pre-collapse magnetic field orientation is
preserved at sites of OH maser emission, as many as one in five
sources could imply a line-of-sight magnetic field opposite to the
assumed underlying regular field.  On the other hand,
\citet{browntaylor} present evidence that the random field may be
partially aligned with the uniform field, rather than being isotropic.
If this is indeed the case, the above estimates of the rate of
discrepancy between the magnetic field direction implied by OH Zeeman
splitting and the regular field direction may be overly pessimistic.
Also, the random component of the Galactic magnetic field may become
less significant between interstellar and GMC densities.

\section{Conclusions and Further Research}

We have found nearly 100 Zeeman pairs of ground-state OH masers using
the data from Paper I.  Combining these Zeeman pairs with others found
in the literature, we have investigated the distribution in the
Galactic plane of the line-of-sight orientation of magnetic fields in
massive SFRs as deduced from OH maser Zeeman splitting.  Our data do
not support the original hypothesis of \citet{davies} that the
large-scale magnetic field traced by OH masers is uniform in the
clockwise sense, nor do we find that the inferred magnetic fields show
a uniform direction in each spiral arm.

All six sources in the second and third Galactic quadrants, which are
typically 1 or 2~kpc distant, have a magnetic field oriented in the
clockwise sense as viewed from the North Galactic Pole, while nearly
all nearby sources in the first and fourth quadrants have a magnetic
field oriented in the counterclockwise direction.  This supports the
hypothesis, deduced from pulsar rotation measures, that a reversal of
the Galactic-scale magnetic field occurs near the galactocentric
radius of the Sun.

Our method for identifying Zeeman pairs in VLA maps of OH masers is
limited by the blending of maser components inevitable with a large
($1\arcsec - 2\arcsec$) beamsize.  The higher resolution of VLBI
allows many more maser spots to be detected and Zeeman pairs to be
identified unambiguously.  The increased detail in VLBI images also
allows for the identification of intra-source magnetic field
reversals, the interpretation of which may be important both for
understanding the conditions of the magnetic field in massive SFRs and
for obtaining enough data to rigorously test the hypothesis that the
magnetic field deduced from OH maser Zeeman splitting in massive SFRs
is correlated with the Galactic magnetic field.  To this end, we are
in the process of imaging interstellar OH masers with the VLBA.  The
results will be published in a future paper.

\begin{deluxetable}{rrr@{/}lr@{/}lr@{/}lr@{/}lr@{/}lr@{/}l}
 \tablecaption{Correlation Results \label{corr_tab}}
% \tablewidth{5.3 truein}
 \tablehead{
% Row 1
   \multicolumn{1}{c}{Separation} & \multicolumn{1}{c}{Pairs of} & 
   \multicolumn{12}{c}{Percentage of Trial Runs Less/More Correlated}\\
% Row 2
   \multicolumn{1}{c}{(kpc)} &    \multicolumn{1}{c}{Sources} & 
   \multicolumn{2}{c}{Overall} & 
   \multicolumn{2}{c}{Car--Sgr} & 
   \multicolumn{2}{c}{Cru--Sct} & \multicolumn{2}{c}{Local} & 
   \multicolumn{2}{c}{Norma} & \multicolumn{2}{c}{Perseus}
  }
  \startdata
% 0.5 &  41&46 & 14&65 & 87&7  &\nodata&\nodata& 25&25 &  0&63 \\
% 1.0 &  31&62 & 63&27 & 14&72 & 13&38 & 50&13 &  0&63 \\
% 2.0 &  90&9  & 87&10 & 76&19 & 44&25 & 97&1  & 19&44 \\
% 5.0 &  51&46 & 52&40 & 66&28 &  0&38 & 99&0  & 45&17 \\
%10.0 &  45&52 & 44&46 & 37&33 &  0&38 & 96&0  & 30&45 \\
%20.0 &  64&35 &  0&68 & 37&33 &  0&38 & 96&0  & 49&18 \\
 0.5 &   43 & 41&46 & 14&65 & 87&7  &\nodata&\nodata& 25&25 &  0&63 \\
 1.0 &  111 & 23&70 & 56&31 & 14&72 &     13&38     & 56&19 &  0&63 \\
 2.0 &  324 & 82&15 & 81&15 & 76&19 &     44&25     & 91&4  & 19&44 \\
 5.0 & 1201 & 52&45 & 56&36 & 66&28 &      0&38     & 95&2  & 45&17 \\
10.0 & 2197 & 37&59 & 42&48 & 37&33 &      0&38     & 89&2  & 30&45 \\
20.0 & 2664 & 63&36 &  0&83 & 37&33 &      0&38     & 89&2  & 49&18 \\
\hline
%\multicolumn{13}{c}{\hrulefill} \\
\multicolumn{2}{l}{Number of Sources}
        & \multicolumn{2}{c}{74} & \multicolumn{2}{c}{22} &
          \multicolumn{2}{c}{17} & \multicolumn{2}{c}{5}  &
          \multicolumn{2}{c}{10} & \multicolumn{2}{c}{9}
  \enddata
\tablecomments{The figures represent the percentage of trial runs for
which the correlation function evaluated on the random data were less
and more correlated than the actual data.  Results to any separation
are cumulative; i.e., all pairs of sources with separation less than
or equal to a given distance are included in the correlation function
results to that distance.  The total number of pairs of distinct
sources with separation less than the figure in the first column is
given in the second column.}
\end{deluxetable}

\begin{deluxetable}{r@{}lr@{.}lccl}
 \tabletypesize{\scriptsize}
 \tablewidth{240pt}
 \tablecaption{Sources and Field Directions by Arm\label{arm_tab}}
 \tablehead{
    \multicolumn{2}{c}{Galactic} & 
       \multicolumn{2}{c}{Galactic} & 
       \colhead{Distance} &
       \colhead{Magnetic} & \colhead{\phantom{References}} \\
    \multicolumn{2}{c}{Longitude} & 
       \multicolumn{2}{c}{Latitude} & 
       \colhead{(kpc)} &
       \colhead{Field\tnm{a}} & \colhead{References}
 }\startdata
\multicolumn{7}{c}{Carina--Sagittarius}\\ \hline
 17.&639 &$+0$&155 & 2.1    & $-$ & 1  \\
 20.&081 &$-0$&135 &12.3\phn& $+$ & 1,7  \\
 32.&744 &$-0$&076 & 2.3    & $-$ & 1,2  \\ 
 34.&258 &$+0$&153 & 3.8    & $+$ & 1,2,3  \\
 35.&024 &$+0$&350 & 3.1    & $+$ & 1,2  \\
 35.&197 &$-0$&743 & 2.2    & $+$ & 1  \\
 35.&577 &$-0$&029 &10.5\phn& $-$ & 1  \\
 43.&796 &$-0$&127 & 3.0    & $+$ & 1,2  \\
 45.&071 &$+0$&134 & 4.7    & $+$ & 1,4,5  \\
 45.&122 &$+0$&133 & 6.0    & $-$ & 1,2,4  \\
 45.&465 &$+0$&060 & 6.0    & $+$ & 1  \\
 49.&469 &$-0$&370 & 6.0    & $+$ & 1,6  \\
 49.&488 &$-0$&387 & 5.5    & $+$ & 1,2,6  \\
%(Flipped +/- in 3rd/4th quadrants)
285.&26  &$-0$&05  & 5.7    & $-$ & 2  \\
294.&51  &$-1$&62  & 1.4    & $-$ & 2  \\
311.&64  &$-0$&38  &14.1\phn& $-$ & 2  \\
344.&581 &$-0$&022 & 2.0    & $-$ & 1  \\
345.&505 &$+0$&347 & 2.1    & $-$ & 1,2  \\
345.&699 &$-0$&090 & 1.6    & $-$ & 1,2  \\
348.&549 &$-0$&978 & 2.1    & $+$ & 1,2,7  \\
351.&161 &$+0$&697 & 2.3    & $-$ & 1  \\
351.&416 &$+0$&646 & 2.0    & $+$ & 1,2,7  \\
\hline
\multicolumn{7}{c}{Crux--Scutum}\\ \hline
  5.&886 &$-0$&393 & 3.8    & $+$ & 1  \\
 11.&03  &$+0$&06  & 3.2    & $-$ & 2  \\
 12.&216 &$-0$&117 & 3.3    & $-$ & 1  \\
 12.&680 &$-0$&181 & 4.0    & $-$ & 1,8  \\
 12.&908 &$-0$&259 & 4.0    & $+$ & 1  \\
 28.&199 &$-0$&048 & 5.7    & $+$ & 1,3  \\ 
 30.&70  &$-0$&06  & 5.5    & $+$ & 7  \\
323.&459 &$-0$&079 & 4.1    & $-$ & 2,10 \\
329.&41  &$-0$&46  & 4.3    & $-$ & 2  \\
331.&34  &$-0$&34  & 4.7    & $-$ & 7  \\
333.&60  &$-0$&21  & 3.5    & $+$ & 2  \\
339.&62  &$-0$&12  & 2.8    & $+$ & 2  \\
341.&219 &$-0$&212 & 3.2    & $-$ & 1  \\
343.&128 &$-0$&063 & 3.1    & $-$ & 1  \\
345.&003 &$-0$&224 & 2.9    & $-$ & 1,2,7  \\
350.&011 &$-1$&341 & 3.1    & $-$ & 1  \\
359.&138 &$+0$&032 & 3.1    & $+$ & 1  \\
\hline
\multicolumn{7}{c}{Local}\\ \hline
 69.&540 &$-0$&976 & 1.3    & $-$ & 1,3,9,11 \\
 81.&77  &$+0$&60  & 2.0    & $-$ & 3  \\
106.&80  &$+5$&31  & 1.4    & $+$ & 3  \\
109.&871 &$+2$&114 & 0.7    & $+$ & 1,3,12 \\
213.&706&$-12$&60  & 0.9    & $+$ & 1  \\
\hline
\multicolumn{7}{c}{Norma}\\ \hline
  8.&67  &$-0$&36  & 4.8    & $+$ & 2  \\
  9.&622 &$+0$&195 & 5.7    & $-$ & 1  \\
 10.&624 &$-0$&385 & 4.8    & $-$ & 1  \\
330.&95  &$-0$&18  & 5.7    & $+$ & 2,7  \\
336.&36  &$-0$&14  &10.3\phn& $+$ & 2  \\
336.&83  &$+0$&02  &10.3\phn& $+$ & 2  \\
337.&61  &$-0$&06  &12.8\phn& $+$ & 2  \\
337.&71  &$-0$&05  &12.2\phn& $+$ & 2,7  \\
344.&42  &$+0$&05  &13.4\phn& $+$ & 2  \\
350.&11  &$+0$&09  & 5.8    & $+$ & 2  \\
\hline    
\multicolumn{7}{c}{Perseus}\\ \hline
 43.&148 &$+0$&015 &11.5\phn& $+$ & 1,2 \\
 43.&165 &$-0$&028 &11.5\phn& $+$ & 1 \\
 43.&167 &$+0$&010 &11.5\phn& $-$ & 7 \\
 70.&293 &$+1$&601 & 8.3    & $-$ & 1,3,11 \\
 75.&782 &$+0$&343 & 5.0    & $+$ & 11 \\
 80.&87  &$+0$&42  & 4.8    & $-$ & 3  \\
111.&533 &$+0$&757 & 2.8    & $+$ & 1  \\
133.&946 &$+1$&064 & 2.2    & $+$ & 1,3,7,13 \\
196.&454 &$-1$&677 & 2.8    & $+$ & 11 \\
\hline
\multicolumn{7}{c}{Unassigned\tnm{b}}\\ \hline
  0.&547 &$-0$&852 & 7.3    & $-$ & 1  \\
  0.&666 &$-0$&034 & 8.2    & $+$ & 1  \\
  0.&670 &$-0$&058 & 8.2    & $-$ & 1  \\
  0.&672 &$-0$&031 & 8.2    & $+$ & 1  \\
 40.&42  &$+0$&70  & 1.2    & $-$ & 1,2  \\
300.&97  &$+1$&15  & 4.4    & $+$ & 2,7  \\ 
305.&81  &$-0$&24  & 3.2    & $-$ & 7  \\
347.&628 &$+0$&149 & 9.8    & $+$ & 1,2  \\
354.&73  &$+0$&29  & 8.5    & $-$ & 2  \\
355.&345 &$+0$&146 &23.1\phn& $+$ & 1  \\
359.&436 &$-0$&103 & 8.2    & $+$ & 1  \\
 \enddata
\tnt{a}{Clockwise ($+$) or counterclockwise ($-$) magnetic field, as
viewed from the North Galactic Pole.  Note that ``positive'' magnetic
fields (pointing away from the Sun) are clockwise in the first and
second Galactic quadrants but counterclockwise in the third and fourth
quadrants.}
\tnt{b}{Includes sources in the Galactic center.}
\tablerefs{
  (1) this paper; 
  (2) \citealt{caswellv}; 
  (3) \citealt{baudry}; 
  (4) \citealt{baartcohen}; 
  (5) \citealt{zhengg45};
  (6) \citealt{gaume}; 
  (7) \citealt{reidsilver}; 
  (8) \citealt{zhengw33};
  (9) \citealt{desmurs}; 
 (10) \citealt{caswellr}; 
 (11) in preparation; 
 (12) \citealt{cohen}; 
 (13) \citealt{garcia}
 }
\end{deluxetable}

\begin{figure}
\epsscale{0.8}
\plotone{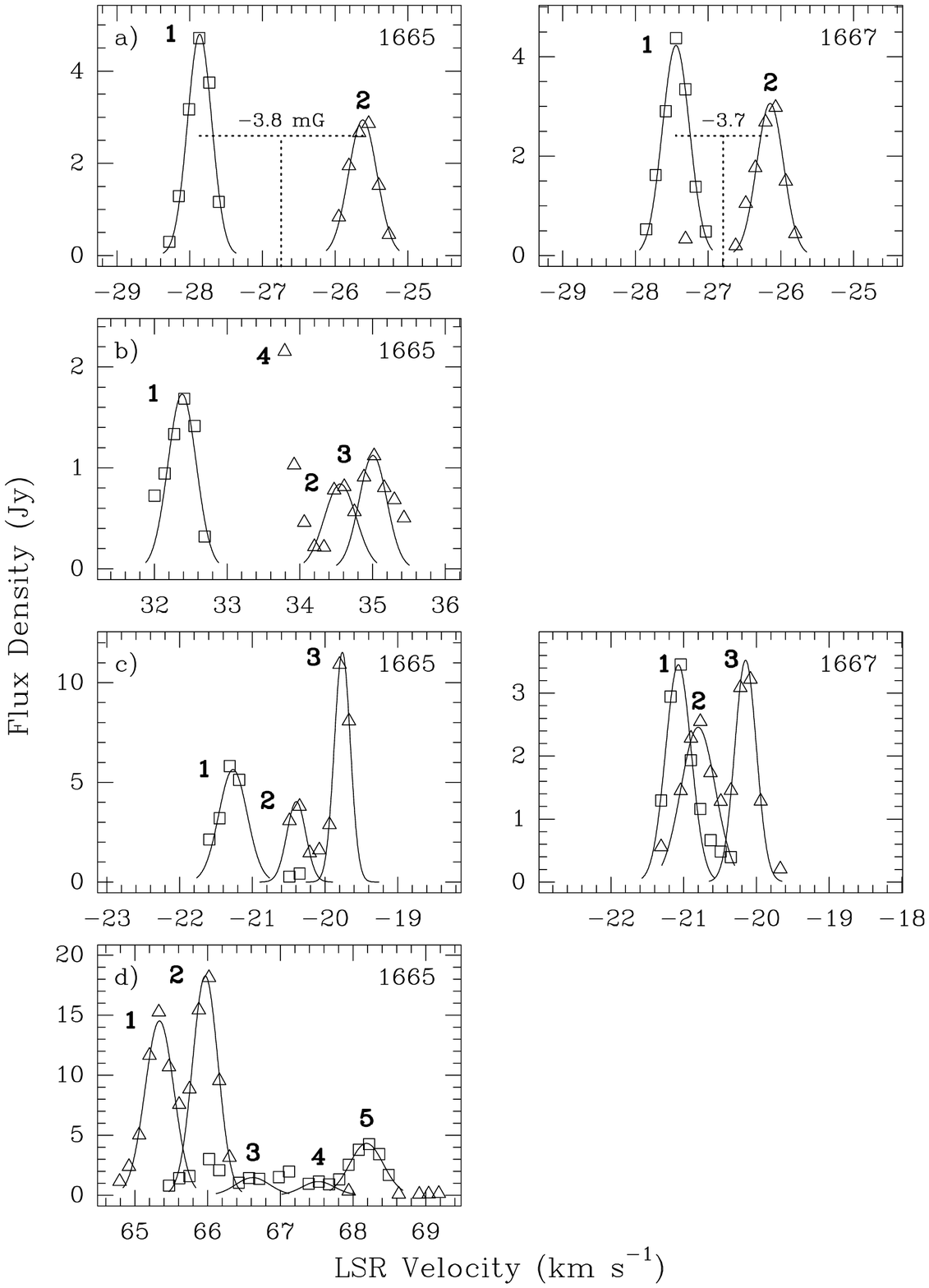}
\caption{\label{samplezeeman}Sample spectra of windows used to find
Zeeman pairs.  Squares and triangles mark RCP and LCP emission,
respectively.  Gaussian fits to the lines are shown, and lines and
features are numbered for reference.  The number in the upper right
corner marks the transition frequency, in MHz.  (a--d) Windows in
G351.775$-$0.538, G032.744$-$0.076, K3$-$50, and G045.465$+$0.047.  The
1665 and 1667~MHz windows in (a) and (c) are centered at approximately
the same position.  The magnetic field and center velocity deduced
from Zeeman splitting are shown in (a).}

\end{figure}

\begin{figure}
\plotone{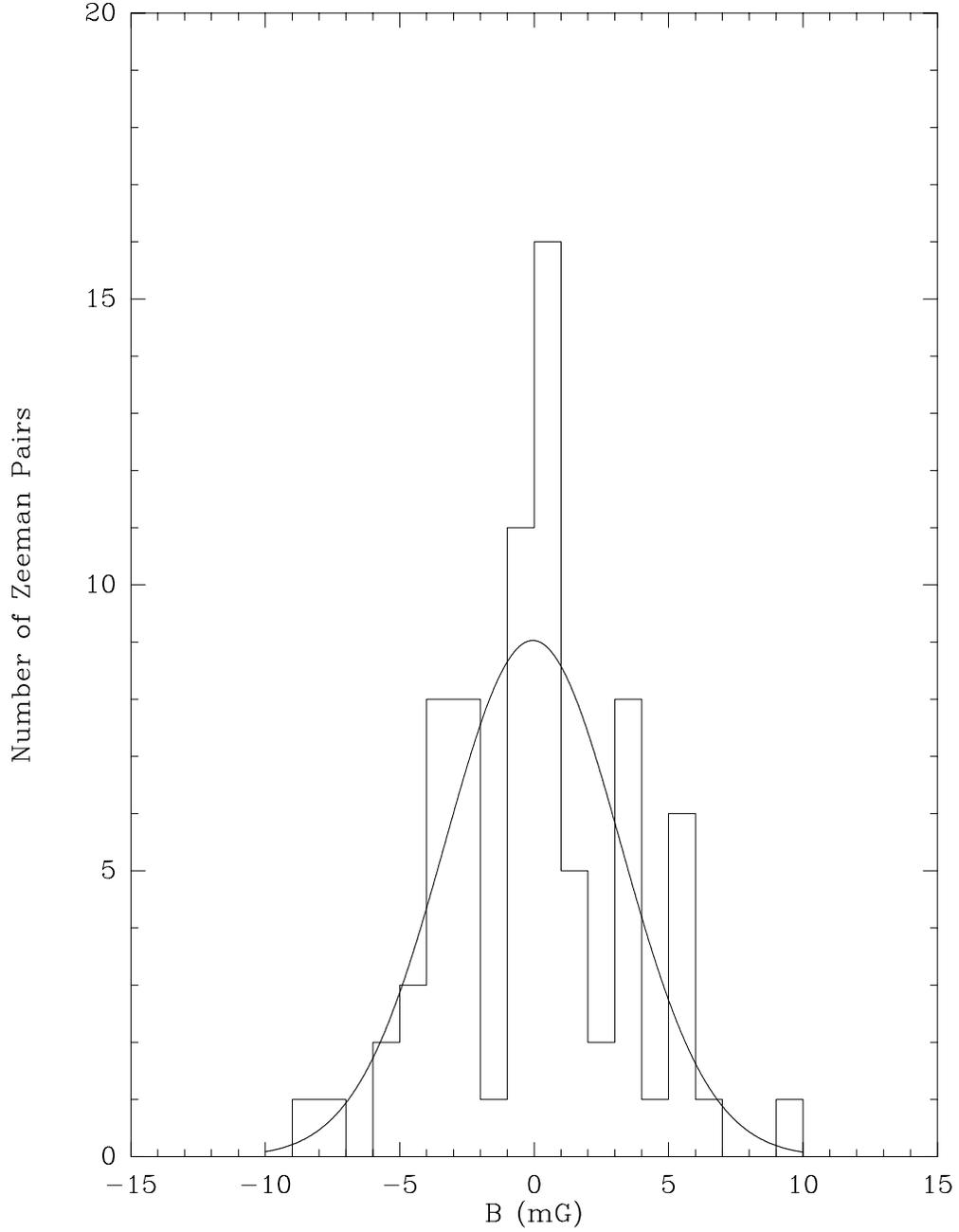}
\caption{\label{histogram}Histogram of the number of Zeeman pairs
found with each magnetic field strength, binned by 1~mG.  Positive
(negative) field strengths represent magnetic fields oriented in the
hemisphere away from (toward) the Sun.  A best-fit Gaussian is also
shown.  The $-25.4$~mG pair in G081.871+0.781 is excluded for reasons
outlined in \S \ref{sourcenotes}.}
\end{figure}

\begin{figure}
\plotone{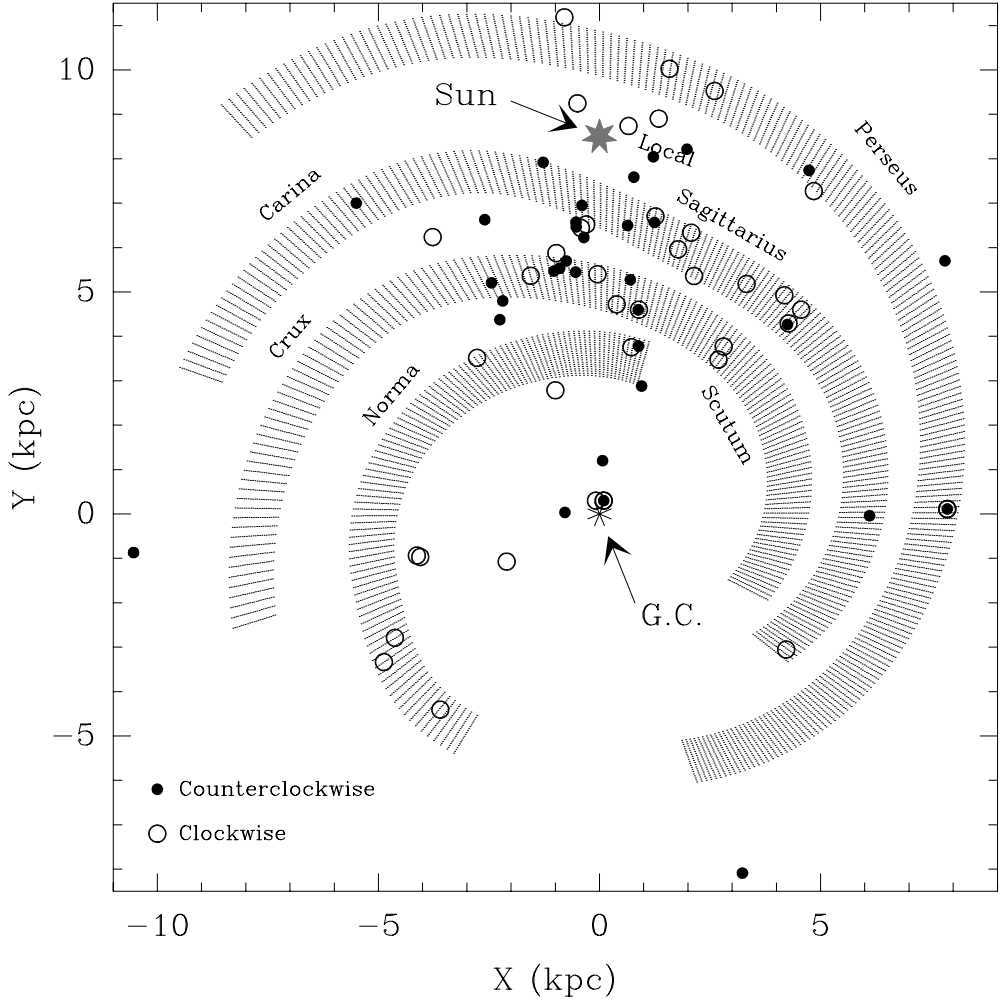}
\caption{\label{galaxy}Line-of-sight magnetic field directions deduced
from OH maser Zeeman splitting.  Seventy-four star-forming regions are
plotted, distinguishing 41 with an overall magnetic field oriented in
a clockwise sense from 33 with field oriented counterclockwise as
viewed from above the Galactic center (i.e., North Galactic Pole).
The locations of the Sun and Galactic center are marked.  The spiral
arms of \citet{taylorcordes} are indicated with shading and labelled
by their common names.}
\end{figure}

\begin{figure}
\plotone{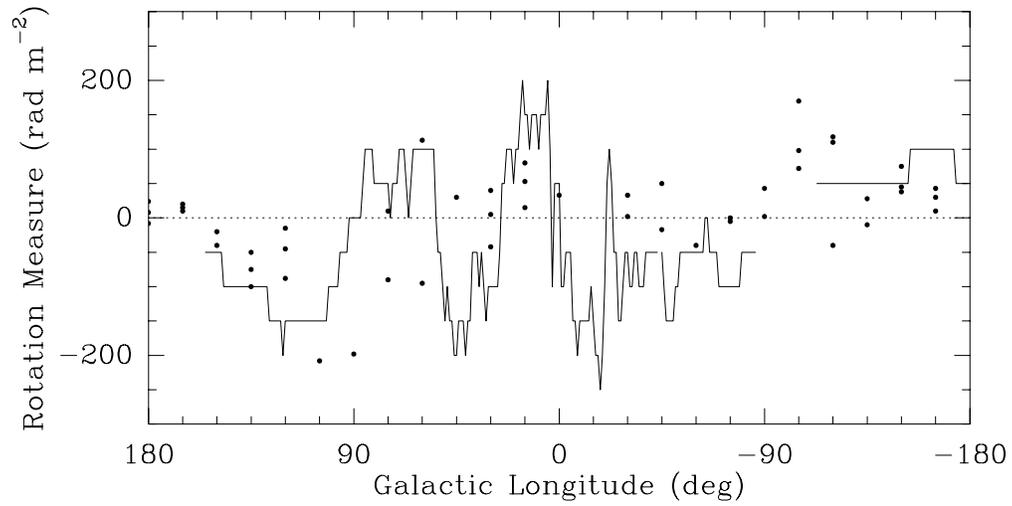}
\caption{\label{rml} Plot of simulated rotation measures generated
from magnetic field directions deduced from OH Zeeman splitting.  No
simulated rotation measures are plotted for Galactic longitudes
unsampled by the data of Figures \ref{galaxy} and \ref{lvplot}.  The
dots represent extragalactic rotation measure data as taken from
\citet{randk} at Galactic latitude $b = -15, 0$, and $15\degr$.  See
\S \ref{rmcomp} for further details.}
\end{figure}

\begin{figure}
\plotone{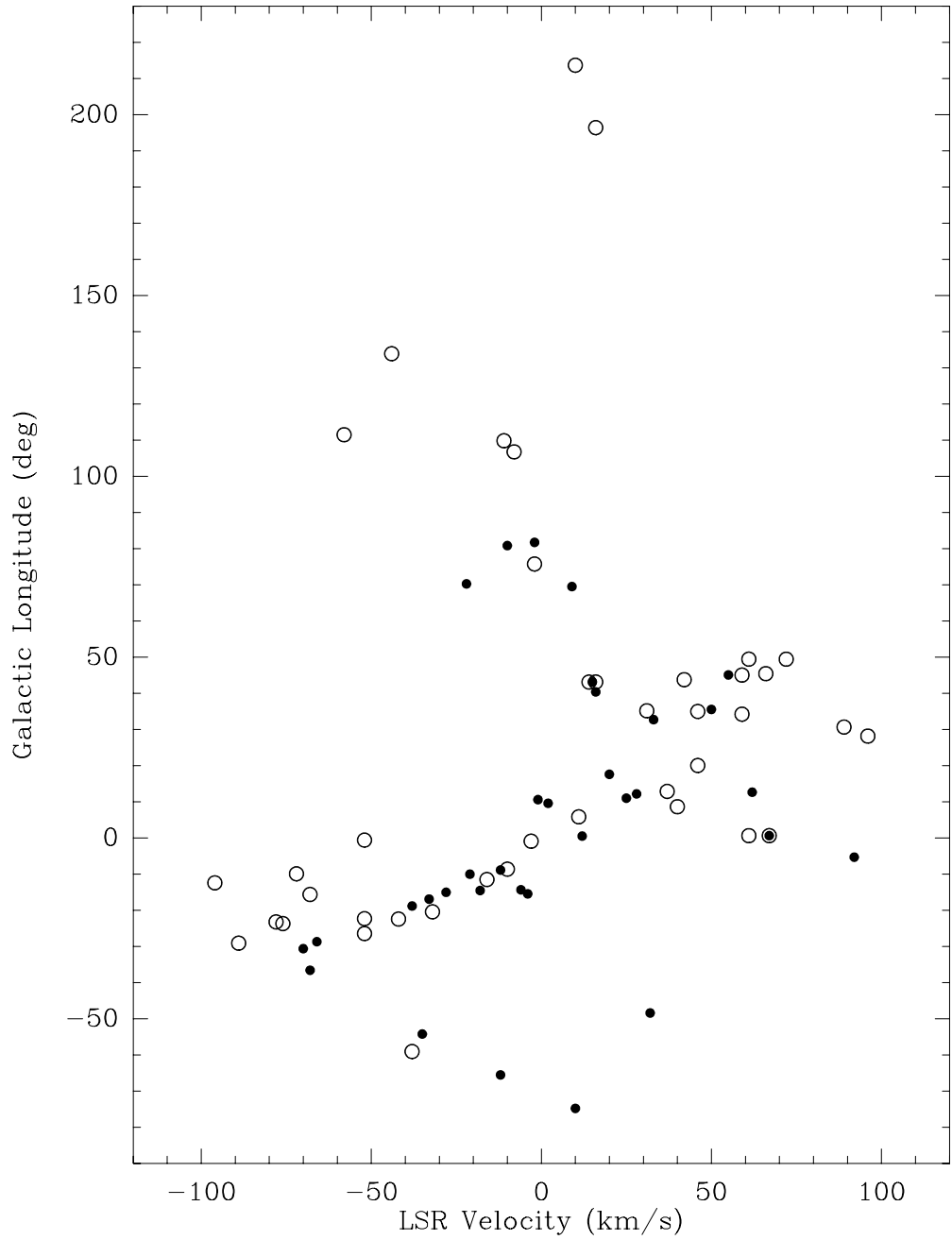}
\caption{\label{lvplot}Longitude-velocity diagram of magnetic field
directions deduced from OH maser Zeeman splitting.  Open and filled
circles are defined as in Figure \ref{galaxy}.}
\end{figure}

\begin{figure}
\plotone{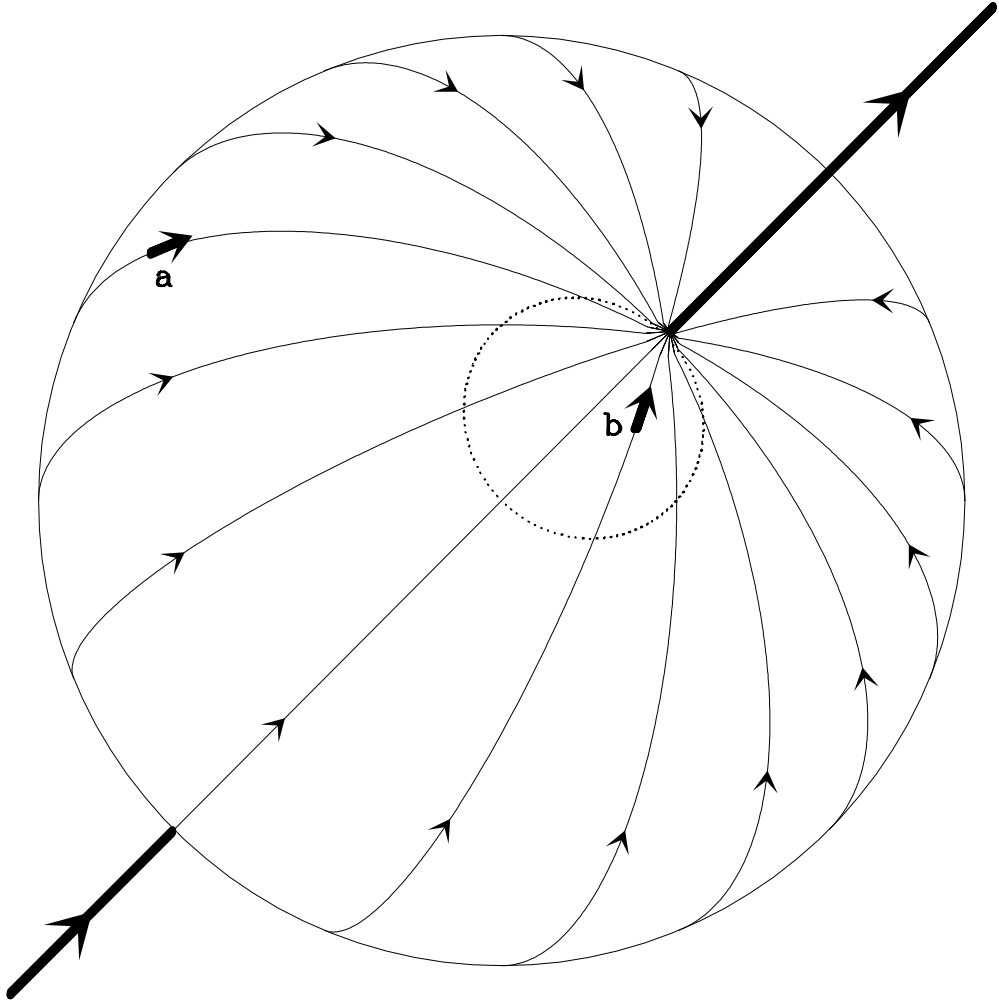}
\caption{\label{sphere}Spherical bubble model of an \ion{H}{2} region
in a magnetic field.  The field is oriented to the upper right and out
of the page, inclined $30\degr$ to the line of sight.  The arrows at
\textbf{a} and \textbf{b} represent the local magnetic field at two
points as might be deduced from Zeeman splitting.  The line-of-sight
projection of the magnetic field is out of the page outside the dotted
curve (as at \textbf{a}) and into the page inside the curve (as at
\textbf{b}).  In the limiting cases of the magnetic field aligned
along (perpendicular to) the line of sight, the line-of-sight
direction of the field is reversed over none (half) of the projected
area of the sphere.  For a uniform distribution of magnetic field
directions intersecting a sphere at all possible angles, on average
the line-of-sight direction of the magnetic field is reversed over
$25\%$ of the area of the projected sphere.}
\end{figure}

\end{document}